\documentclass[twocolumn,floats,floatfix,superscriptaddress,nofootinbib, amssymb, preprintnumbers]{revtex4-1}

\usepackage{graphicx}
\usepackage{bm}
\usepackage{xcolor}
\usepackage{booktabs}
\PassOptionsToPackage{breaklinks}{hyperref}
\usepackage[linktocpage,colorlinks=true, linkcolor=bluscuro,urlcolor=bluscuro,citecolor=bluscuro]{hyperref}
\definecolor{bluscuro}{rgb}{0.15, 0.2, .85}
\usepackage{amsmath, amssymb}
\usepackage{slashed}
\usepackage{latexsym}
\usepackage{braket}
\usepackage[export]{adjustbox}
\usepackage{color}
\usepackage{multirow}
\usepackage{xspace}
\usepackage{comment}
\usepackage[caption=false]{subfig}
\usepackage{cancel}
\usepackage{url}
\usepackage{nicematrix}
\usepackage{needspace}

\newcommand{\sapienzainfn}{Dipartimento di Fisica, Sapienza Università 
	di Roma \& INFN, Sezione di Roma, Piazzale Aldo Moro 5, 00185, Roma, Italy}

\begin{document}

\title{A Multipole-Based Framework for Kerr Black Hole Mimickers:\\ From General Construction to a Specific Case of Study}

\author{Claudio Gambino}
\email{claudio.gambino@uniroma1.it}
\affiliation{\sapienzainfn}

\begin{abstract}
We present a novel systematic framework for engineering energy-momentum tensors in linearized gravity that generate gravitational fields with a prescribed multipolar structure, aimed at modeling black hole mimickers. As a concrete example, we analyze an anisotropic rotating fluid with a Gaussian-like energy-density profile which, at linearized order, satisfies all energy and causality conditions while generating a gravitational field that exactly reproduces the multipolar structure of a Kerr black hole. We then investigate the phenomenology of both the source and the induced metric, comparing our findings with the corresponding black hole limit.
\end{abstract}

\maketitle

\tableofcontents

\needspace{5\baselineskip}
\section{Introduction}\label{sec:Introduction}

Black hole (BH) physics is among the most active and productive areas of research in theoretical and experimental physics. It boasts a long history of foundational developments~\cite{Schwarzschild:1916uq, Kerr:1963ud,Misner:1973prb}, alongside more recent experimental breakthroughs such as gravitational-wave detections~\cite{LIGOScientific:2016aoc, LIGOScientific:2019fpa} and BH imaging~\cite{EventHorizonTelescope:2019dse, EventHorizonTelescope:2022xqj}. The peculiarity of BHs among other compact objects stems from several key features, including their thermodynamic properties~\cite{Bardeen:1973gs, Hawking:1974rv}, the no-hair theorem~\cite{Israel:1967wq, Carter:1971zc}, and the presence of event horizons and curvature singularities. In particular, classical BH solutions, such as Schwarzschild and Kerr, possess an event horizon (with both outer and inner horizons in the Kerr case), where light experiences infinite gravitational redshift for a distant observer, as well as a central singularity where spacetime curvature diverges.

These extreme properties suggest that our current understanding of gravity may be incomplete in high-curvature regimes, and that a quantum theory of gravity could predict objects that resemble BHs macroscopically but differ at short distances~\cite{Giddings:1992hh, Lunin:2001jy,Mathur:2005zp,Hayward:2005gi, Bena:2005va, Bena:2006kb, Mathur:2008nj,Frolov:2016pav, Cano:2018aod, Carballo-Rubio:2018pmi, Simpson:2018tsi, Simpson:2019cer, Bianchi:2020bxa, Bianchi:2020miz} (see also~\cite{Buoninfante:2024oxl, Carballo-Rubio:2025fnc} for recent reviews). This perspective has sparked growing interest in BH mimickers: compact objects that reproduce some macroscopic BH properties while deviating in specific limits~\cite{Cardoso:2017cqb, Mark:2017dnq, Carballo-Rubio:2018jzw, Mazza:2021rgq, Cardoso:2022fbq, Casadio:2024lgw}. Beyond their theoretical appeal, such models can also be probed observationally, potentially offering windows into new physics~\cite{Cardoso:2019rvt, Abedi:2016hgu, Jiang:2021ajk, Shaikh:2022ivr} (see~\cite{Bambi:2025wjx} for a recent review).

A compelling strategy for constructing BH mimickers is to impose that they share the same gravitational multipolar structure as classical BHs~\cite{Ryan:1995wh, Pappas:2012ns}. While the Schwarzschild geometry is fully described by a single monopole moment (the mass), Kerr BHs possess an infinite tower of mass and current multipole moments uniquely fixed by their mass and angular momentum~\mbox{\cite{Geroch:1970cc, Geroch:1970cd, Hansen:1974zz, Thorne:1980ru, Gursel1983}}. However, much like in Newtonian gravity, a given set of multipole moments does not uniquely determine the matter source. Therefore, the observation of a gravitational field with the Kerr multipolar structure does not guarantee the presence of an actual Kerr BH. Furthermore, it remains an open question whether stable, physically viable matter configurations can exactly approach the Kerr geometry at infinity~\cite{Friedman:1978ygc,Cardoso:2007az,Moschidis:2016zjy}.

In fact, while spherically symmetric objects generically approach Schwarzschild at large distances, stationary spacetimes do not necessarily converge to a unique asymptotic form. The BH uniqueness theorems establish Kerr as the only stationary, axisymmetric vacuum BH solution in four-dimensional General Relativity (GR)~\cite{Israel:1967wq, Hawking:1971vc,Carter:1971zc, Hawking:1973uf, Robinson:1975bv}, but these theorems can be evaded by considering either non-vacuum spacetimes or horizonless configurations—or both. In this context, identifying sources that exactly reproduce the asymptotic structure of Kerr is a nontrivial task, and a systematic study could shed light on the scope and limitations of uniqueness theorems, particularly in higher dimensions where no uniqueness theorem (in a strict sense) is known~\cite{Emparan:2025wsh}. It may also offer new insight into the connection between Kerr uniqueness and its multipolar content.

The aim of this paper then is to take a step forward in this direction by investigating whether a stable, rotating matter configuration can (\textit{i}) source a gravitational field that precisely matches the Kerr geometry at large distances, thus reproducing its entire multipolar structure, while (\textit{ii}) satisfying energy and causality conditions.

To this end, we build on the momentum-space formalism developed in~\cite{Bianchi:2024shc}, where it was shown that, order by order in the spin expansion, the Fourier transform of a linearized energy-momentum tensor (EMT) naturally organizes into a series of \textit{form factors}. These form factors characterize the gravitational multipoles of the induced spacetime and offer a relativistic generalization of the Newtonian notion of source multipoles, generalizing the post-Newtonian analysis of~\cite{Bonga:2021ouq}. However, specifying the form factors fixes only the asymptotic multipolar structure—not the full local profile of the source. As a result, for any given multipole spectrum, an infinite equivalence class of physically distinct EMTs can be constructed.

These differences are encoded in \textit{structure functions}—analytic functions of the transferred momentum that capture the internal structure of the source. By analogy with particle physics, setting structure functions to unity corresponds to describing a point-like object. Indeed, in~\cite{Bianchi:2024shc} it was shown that imposing Kerr multipoles with identity structure functions leads to a singular ring-like source that generates the linearized Kerr metric. In this work, we go beyond that construction by introducing a non-trivial structure function that smears the singularity, yielding a regular EMT whose associated gravitational field still asymptotically matches that of a Kerr BH.

As a specific case of study, we consider a Gaussian structure function and construct the associated EMT. This describes an anisotropic, rotating fluid that sources a Kerr mimicker and satisfies the energy and causality conditions at linearized order. The source is characterized by its mass, angular momentum, and a set of characteristic length scales ${R_n}$, which must be suitably constrained for the EMT to remain real-valued. We identify a region of parameter space where this condition is satisfied and all physical requirements hold. In this setup, we analyze both the EMT and the corresponding linearized metric in harmonic gauge, demonstrating explicitly how the curvature singularity is resolved into a smooth distribution over a finite region, while the gravitational field retains the full Kerr multipolar structure at large distances.

Although our construction is exact in the angular momentum expansion, it remains perturbative in the gravitational coupling. Indeed, the proposed formalism captures the full multipolar structure of the spacetime but only at linear order in gravitational coupling constant ${G_N}$, and does not account for the non-linear gravitational backreaction on the geometry. Extending this framework to fully non-perturbative configurations, where the metric satisfies Einstein’s equations beyond the linearized regime, remains a significant open challenge. Such an extension would be crucial for determining whether regular, horizonless mimickers can exist as exact solutions to GR with realistic matter content.

The paper is organized as follows. In Sec.~\ref{sec:GeneralConstruction}, we review the momentum-space formalism and derive a general expression for the EMT in terms of form factors and structure functions. In Sec.~\ref{sec:GaussianSF}, we specialize to Gaussian structure functions and obtain an explicit expression for the EMT. In Sec.~\ref{sec:KerrCase}, we impose the Kerr multipolar structure, and in Sec.~\ref{sec:EMT_KerrPheno}, we match the resulting EMT to a rotating anisotropic fluid, analyzing the resulting energy-density, pressure, and rotational velocity for different parameter choices. In Sec.~\ref{sec:MetricPheno}, we study the associated linearized metric and compare it with the Kerr case. Finally, Sec.~\ref{sec:Conclusions} summarizes our findings and provides an outlook on future prospects.

\textbf{Conventions.}
We work in mostly plus signature, ${\eta_{00} = -1}$, and use natural units ${\hbar = c = 1}$ for analytic computations, keeping the gravitational coupling constant ${G_N}$ explicit. For numerical plots, we further set $G_N = m = 1$, so that all plotted quantities are dimensionless. Greek indices ${\mu, \nu = 0, 1, ..., d}$ denote spacetime components, while Latin indices ${i, j = 1, ..., d}$ refer to spatial ones, where the number of spacetime dimensions is ${D = d+1}$. Notice that while some intermediate results are formulated in generic ${D}$ dimensions for generality, we restrict to ${D = 4}$ in all explicit computations and physical interpretations.

\needspace{5\baselineskip}
\section{General construction}\label{sec:GeneralConstruction}

Let us consider GR in arbitrary spacetime dimensions, where a generic stationary metric ${g_{\mu\nu}}$ is linearized around a flat background ${\eta_{\mu\nu}}$ as
\begin{equation}
    g_{\mu\nu} = \eta_{\mu\nu} + h_{\mu\nu} + O(G_N^2)\ ,
\end{equation}
where ${h_{\mu\nu}}$ is the metric perturbation at first order in the gravitational coupling constant ${G_N}$.
By solving the linearized Einstein equations
\begin{equation}
    \Box h_{\mu\nu}(\vec x) = 16 \pi G_N P_{\mu\nu, \rho\sigma} T^{\rho\sigma}(\vec x)\ ,
\end{equation}
where ${\Box = \eta^{\mu\nu} \partial_\mu \partial_\nu}$, ${P_{\mu\nu, \rho\sigma}}$ is the transverse projector of the graviton propagator in a given gauge, and ${T_{\mu\nu}}$ is the linearized EMT sourcing the gravitational field, one obtains
\begin{equation}\label{eq:FormalMetricfromEMT}
    h_{\mu\nu}(\vec x) = 16\pi G_N \int \frac{d^d \vec q}{(2\pi)^d} \frac{e^{-i \vec q \cdot \vec x}}{\vec q\, ^2} P_{\mu\nu, \rho \sigma} T^{\rho\sigma}(\vec q\, )\ ,
\end{equation}
with ${T_{\mu\nu}(\vec q\, )}$ the Fourier transform of the EMT, defined as
\begin{equation}
    T_{\mu\nu}(\vec q\, ) = \int d^d \vec x\, e^{i \vec q \cdot \vec x} T_{\mu\nu}(\vec x)\ ,
\end{equation}
and ${d}$ the number of spatial dimensions. Since we are considering a stationary spacetime, both the metric perturbation and the EMT are time-independent, hence ${q_0=0}$ and so ${q^2=\vec q\, ^2}$. We will keep this understood from now on, and write for instance ${T(\vec q\, ) \equiv T(q)}$.

Our goal is to relate local properties of the EMT in momentum space to the long-range behavior of the linearized metric. To this end, we briefly review the momentum-space formalism in GR introduced in~\cite{Bianchi:2024shc}. Consider a stationary source described only by mass and spin, with ${u^\mu}$ the four-velocity of its center of mass and ${S^{\mu\nu}}$ its spin density tensor, satisfying ${S^{\mu\nu} u_\nu = q^\mu u_\mu = 0}$. The EMT in momentum space can be written in terms of so-called \textit{form factors} as
\begin{equation}\label{eq:EMTdef}
\begin{aligned}
    &T^{\mu\nu}(q) = m\, u^\mu u^\nu \sum_{\ell=0}^{+\infty} F_{2\ell, 1}\, \zeta^{2\ell}\\
    &+ m\, S^{\mu\sigma} q_\sigma\, S^{\nu\lambda} q_\lambda \sum_{\ell=0}^{+\infty} F_{2\ell+2, 2}\, \zeta^{2\ell} \\
    & + \frac{i}{2} m \left( u^\mu S^{\nu\sigma} q_\sigma + u^\nu S^{\mu\sigma} q_\sigma \right) \sum_{\ell=0}^{+\infty} F_{2\ell+1, 3}\, \zeta^{2\ell}\ ,
\end{aligned}
\end{equation}
where
\begin{equation}
    \zeta = \sqrt{-q^\mu S_{\mu}{}^{\nu} S_{\nu}{}^{\sigma} q_{\sigma}}\ ,
\end{equation}
and the ${F_{\ell, n}}$ are constant form factors, with ${F_{0,1} = F_{1,3} = 1}$ fixing the ADM mass and angular momentum, and ${F_{0,2} = 0}$ by definition.

As shown in~\cite{Bianchi:2024shc}, the form factors, which parametrize the linearized source, are in one-to-one correspondence with the gravitational multipoles of the induced spacetime. Consider in fact the multipole expansion of a linearized metric in ACMC coordinates~\cite{Thorne:1980ru} within the generalized Thorne formalism
\begin{equation}\label{eq:GeneralizedTFE}
\begin{aligned}
    g_{00} &= -1 + 4\frac{d-2}{d-1} \sum_{\ell=0}^{+\infty} \frac{G m\, \rho(r)}{r^\ell} \mathbb{M}^{(\ell)}_{A_\ell} N_{A_\ell} + \cdots\ , \\
    g_{0i} &= 2(d-2) \sum_{\ell=1}^{+\infty} \frac{G m\, \rho(r)}{r^\ell} \mathbb{J}^{(\ell)}_{i, A_\ell} N_{A_\ell} + \cdots\ , \\
    g_{ij} &= \delta_{ij} + 4\frac{d-2}{d-1} \sum_{\ell=2}^{+\infty} \frac{G m\, \rho(r)}{r^\ell} \tilde{\mathbb{G}}^{(\ell)}_{ij, A_\ell} N_{A_\ell} + \cdots\ ,
\end{aligned}
\end{equation}
where ${A_\ell = a_1 \cdots a_\ell}$ and ${N_{A_\ell} = x_{a_1} \cdots x_{a_\ell} / r^\ell}$, with ${r^2 = x_1^2 + \cdots + x_d^2}$.

In Eq.~\eqref{eq:GeneralizedTFE}, the tensors ${\mathbb{M}^{(\ell)}_{A_\ell}}$ and ${\mathbb{J}^{(\ell)}_{i, A_\ell}}$ are the mass and current multipole moments, while ${\tilde{\mathbb{G}}^{(\ell)}_{ij, A_\ell}}$ are related to the stress multipoles by
\begin{equation}
    \mathbb{G}^{(\ell)}_{ij, A_\ell} = \tilde{\mathbb{G}}^{(\ell)}_{ij, A_\ell} + \frac{1}{2} \delta_{ij} \left( \mathbb{M}^{(\ell)}_{A_\ell} - \tilde{\mathbb{G}}^{(\ell)}_{kk, A_\ell} \right)\ ,
\end{equation}
with ${\mathbb{G}^{(\ell)}_{ij, A_\ell}}$ defined in~\cite{Gambino:2024uge} (see also~\cite{Amalberti:2023ohj}).

One can express the gravitational multipoles in terms of form factors, interpreting them as \textit{source multipoles} in analogy with Newtonian gravity as in 
\begin{widetext}
\begin{equation}\label{eq:GravitationalMultipoles}
\begin{aligned}
    \mathbb{M}^{(2\ell)}_{A_{2\ell}} &= \frac{(d+4\ell-4)!!}{(d-2)!!} (-1)^\ell \left( F_{2\ell, 2} + (d-2) F_{2\ell, 1} \right) (-S \cdot S)_{A_{2\ell}} \Big|_{\rm STF}\ , \\
    \mathbb{J}^{(2\ell+1)}_{i, A_{2\ell+1}} &= \frac{(d+4\ell-2)!!}{(d-2)!!} (-1)^\ell F_{2\ell+1, 3} \, S_{ia_1} (-S \cdot S)_{A_{2\ell}} \Big|_{\rm ASTF}\ , \\
    \mathbb{G}^{(2\ell)}_{ij, A_{2\ell}} &= (d-1)\frac{(d+4\ell-4)!!}{(d-2)!!} (-1)^\ell F_{2\ell, 2} \, S_{ia_1} S_{ja_2} (-S \cdot S)_{A_{2\ell-2}} \Big|_{\rm RSTF}\ ,
\end{aligned}
\end{equation}
\end{widetext}
with
\begin{equation}
    \mathbb{M}^{(2\ell+1)}_{A_{2\ell+1}} = 0\ ,\quad
    \mathbb{J}^{(2\ell)}_{i, A_{2\ell}} = 0\ ,\quad
    \mathbb{G}^{(2\ell+1)}_{ij, A_{2\ell+1}} = 0\ ,
\end{equation}
for every ${\ell = 0, 1, 2, \dots}$, and using the shorthand notation
\begin{equation}
    (S \cdot S)^{A_{2n}} \equiv S^{a_1 \lambda} S_{\lambda}{}^{a_2} \cdots S^{a_{n-1} \sigma} S_\sigma{}^{a_n}\ .
\end{equation}
Notice that each multipole tensor in Eq.~\eqref{eq:GravitationalMultipoles} is symmetrized according to its index type. Although this symmetrization is not needed when contracted with ${N_{A_\ell}}$, it remains important in the general classification. Here STF denotes totally symmetric and trace-free tensors~\cite{Thorne:1980ru}, while ASTF and RSTF denote irreducible representations of ${SO(d)}$ with one or two antisymmetrizations~\cite{Heynen:2023sin, Gambino:2024uge}.

As in Newtonian gravity, fixing the multipoles does not uniquely determine the source. In the linearized regime, this ambiguity arises in two ways: either by modifying the tensorial structure of the EMT in Eq.~\eqref{eq:EMTdef} through the addition of terms that do not contribute asymptotically (see~\cite{Bianchi:2024shc}), or by altering the local structure of the source through momentum-dependent functions. Focusing on the latter, one observes that constant form factors describe ``point-like'' sources, and it is possible to promote them to analytic functions of ${q^2}$, thereby giving the source a non-trivial structure without modifying its multipolar content. This is because any term proportional to ${q^2}$ cancels the graviton propagator in Eq.~\eqref{eq:FormalMetricfromEMT}, leading to contributions in the gravitational field that are localized (\textit{i.e.} delta functions and their derivatives).

To construct such an equivalence class explicitly, we now restrict to ${d = 3}$ spatial dimensions, although the method generalizes to higher-dimensional spacetimes. Let us consider Cartesian coordinates ${\vec{x} = (x, y, z)}$ and move to a frame in which the spin is aligned along the ${z}$-axis. In this frame, the spin tensor reads
\begin{equation}
S^{ij} = 
\begin{pNiceMatrix}[columns-width=auto]
0 & a & 0 \\
-a & 0 & 0 \\
0 & 0 & 0
\end{pNiceMatrix}\ ,
\end{equation}
where ${a}$ is the angular momentum density, defined as ${a = J/m}$ in terms of the ADM angular momentum ${J}$ and mass ${m}$. Since we are working in four spacetime dimensions, one has ${\zeta = a q_\perp}$, where ${q_\perp^2 = q_x^2 + q_y^2}$.

We can now formally resum the form factor expansion into
\begin{equation}
F_n(a q_\perp) = \sum_{\ell=0}^{+\infty} F_{\ell, n} (a^2 q_\perp^2)^\ell\ .
\end{equation}
Adding a local structure corresponds to promoting these functions to analytic functions of ${q^2}$,
\begin{equation}\label{eq:promotion}
F_n(a q_\perp) \rightarrow F_n(a q_\perp)\, K_n(q^2)\ ,
\end{equation}
where ${K_n(q^2)}$ are the \textit{structure functions}, such as
\begin{equation}
    K_n(q^2)=1+\sum_{i=1}^{+\infty}a_{i}^{(n)}\, q^{2i}\ .
\end{equation}
Eq.~\eqref{eq:promotion} is only one example of such promotion; more involved constructions are certainly possible and may lead to richer physical scenarios.

Introducing the spin vector ${s^i = (0, 0, a)}$, defined via ${S^{ij} = \varepsilon^{ijk} s_k}$ with $\varepsilon^{ijk}$ the Levi-Civita symbol, we can rewrite the EMT as
\begin{equation}\label{eq:EMTinMomentum}
\begin{aligned}
T^{00}(q) &= m\, F_1(a q_\perp)\, K_1(q^2)\ , \\
T^{ij}(q) &= m\, (s \times q)^i (s \times q)^j\, F_2(a q_\perp)\, K_2(q^2)\ , \\
T^{0i}(q) &= -\frac{i}{2} m\, (s \times q)^i\, F_3(a q_\perp)\, K_3(q^2)\ .
\end{aligned}
\end{equation}
Our goal is now to perform the Fourier transform of Eq.~\eqref{eq:EMTinMomentum} and obtain a compact expression for the EMT in coordinate space, valid for arbitrary form factors and structure functions.

Let us first focus on the ${T^{00}}$ component. Using
\begin{equation}
K_1(q^2) = \int d^3 x'\, e^{i q \cdot x'} K_1(r'^2)\ ,
\end{equation}
and substituting it into the definition of ${T^{00}}$, we find
\begin{equation}
T^{00}(\rho, z) = \int \frac{d^3 q}{(2\pi)^3} e^{-i q \cdot x} F_1(a q_\perp) \int d^3 x'\, e^{i q \cdot x'} K_1(\rho'^2 + z'^2)\ .
\end{equation}
Since the integrand does not depend on ${q_z}$, the integration over ${q_z}$ yields a delta function ${\delta(z - z')}$, and the expression becomes
\begin{widetext}
\begin{equation}
T^{00}(\rho, z) = \int_0^{+\infty} dq_\perp\, q_\perp \int_0^{+\infty} d\rho'\, \rho'\, J_0(q_\perp \rho)\, J_0(q_\perp \rho')\, F_1(a q_\perp)\, K_1(\rho'^2 + z^2)\ ,
\end{equation}
where we switched to cylindrical coordinates ${\vec{x} = (\rho, \phi, z)}$.
Expanding the form factor function order by order in angular momentum, we then obtain
\begin{equation}
T^{00}(\rho, z) = \sum_{\ell=0}^{+\infty} F_{2\ell, 1} a^{2\ell} \int_0^{+\infty} dq_\perp\, q_\perp^{2\ell+1} \int_0^{+\infty} d\rho'\, \rho'\, J_0(q_\perp \rho)\, J_0(q_\perp \rho')\, K_1(\rho'^2 + z^2)\ .
\end{equation}
\end{widetext}

Using the identity
\begin{equation}
\frac{1}{\rho} \partial_\rho \Big( \rho\, \partial_\rho J_0(q_\perp \rho) \Big) = \nabla_\rho^2 J_0(q_\perp \rho) = -q_\perp^2 J_0(q_\perp \rho)\ ,
\end{equation}
where ${\nabla_\rho^2}$ is the radial part of the Laplacian in cylindrical coordinates,
\begin{equation}
\nabla^2 = \nabla_\rho^2 + \frac{1}{\rho^2} \frac{\partial^2}{\partial \phi^2} + \frac{\partial^2}{\partial z^2}\ ,
\end{equation}
we can rewrite the expression as
\begin{widetext}
    \begin{equation}
T^{00}(\rho, z) = \sum_{\ell=0}^{+\infty} (-1)^\ell F_{2\ell, 1} a^{2\ell} (\nabla_\rho^2)^\ell \int_0^{+\infty} dq_\perp\, q_\perp \int_0^{+\infty} d\rho'\, \rho'\, J_0(q_\perp \rho)\, J_0(q_\perp \rho')\, K_1(\rho'^2 + z^2)\ .
\end{equation}
\end{widetext}
Then, using the orthogonality of Bessel functions,
\begin{equation}
\int_0^{+\infty} du\, u\, J_0(\alpha u)\, J_0(\beta u) = \frac{\delta(\alpha - \beta)}{\alpha}\ ,
\end{equation}
one finally obtains
\begin{equation}
T^{00}(\rho, z) = \sum_{\ell=0}^{+\infty} (-1)^\ell F_{2\ell, 1} a^{2\ell} (\nabla_\rho^2)^\ell K_1(\rho^2 + z^2)\ .
\end{equation}

Repeating the same argument for the other components, we obtain the full EMT in coordinate space, valid at every order in the spin expansion
\begin{widetext}
\begin{equation}\label{eq:EMTwithRotationFromNR}
\begin{aligned}
T^{00}(\rho, z) &= m \sum_{\ell=0}^{+\infty} (-1)^\ell F_{2\ell, 1} a^{2\ell} (\nabla_\rho^2)^\ell K_1(\rho^2 + z^2)\ , \\
T^{ij}(\rho, z) &= -m\, (s \times \partial)^i (s \times \partial)^j \sum_{\ell=0}^{+\infty} (-1)^\ell F_{2\ell+2, 2} a^{2\ell} (\nabla_\rho^2)^\ell K_2(\rho^2 + z^2)\ , \\
T^{0i}(\rho, z) &= \frac{1}{2} m\, (s \times \partial)^i \sum_{\ell=0}^{+\infty} (-1)^\ell F_{2\ell+1, 3} a^{2\ell} (\nabla_\rho^2)^\ell K_3(\rho^2 + z^2)\ .
\end{aligned}
\end{equation}
\end{widetext}

To interpret the real-space structure functions, consider the non-rotating case ${a = 0}$, where the EMT reduces to ${T_{\mu\nu}(r) = u_\mu u_\nu \epsilon(r)}$, and ${\epsilon(r)=m K_1(r^2)}$ corresponds to the energy density. More generally, the functions ${K_n(r^2)}$ encode physical properties of the energy distribution, such as rotational velocity and pressure. This means that, when restricting to spin-induced multipoles, starting from a spherically symmetric source, Eq.~\eqref{eq:EMTwithRotationFromNR} provides the recipe to make it rotate exactly in the way required to reproduce a specific multipolar structure determined by the form factors.

\needspace{5\baselineskip}
\section{Gaussian structure functions}\label{sec:GaussianSF}

Up to this point, the construction has been entirely general and does not assume any particular form for the multipole coefficients or the structure functions. We now focus on a specific case of study, namely a source with a Gaussian energy-density profile.

Consider a Gaussian structure function of the form
\begin{equation}
    K_n(q^2) = e^{-q^2 R_n^2}\ ,
\end{equation}
which leads to the energy density
\begin{equation}\label{eq:EnergyDensityDef}
    \epsilon(r) = m \int \frac{d^3 q}{(2\pi)^3} e^{-i q \cdot x} K_1(q^2) = \frac{m}{8\pi^{3/2} R_1^3} e^{-\frac{r^2}{4 R_1^2}}\ ,
\end{equation}
where the ${R_n}$'s are new characteristic length scales of the system. The Gaussian energy profile not only smooths out potential singularities, as any analytic structure function does, but also coincides with the type of matter distribution used in noncommutative-inspired models~\cite{Nicolini:2005vd}, where it emerges as an effective way to regularize curvature divergences. Moreover, this choice proves particularly convenient for analytic calculations, as it diagonalizes the power series of the Laplacian operator in Eq.~\eqref{eq:EMTwithRotationFromNR}.

Indeed, one can show that
\begin{equation}\label{eq:EigenValueRelation}
    (\nabla_\rho^2)^\ell \epsilon(r) = \left( \frac{m\, e^{-\frac{z^2}{4 R_1^2}}}{8\pi^{3/2} R_1^3} \right) \frac{(-1)^\ell \ell!}{R_1^{2\ell}}\, {}_1F_1\left(\ell+1, 1, -\frac{\rho^2}{4 R_1^2} \right)\ ,
\end{equation}
where ${ {}_1F_1(a, b, z) }$ is the confluent hypergeometric function (Kummer function).

To prove Eq.~\eqref{eq:EigenValueRelation}, note that the Gaussian can be written in terms of a Kummer function as in
\begin{equation}
    \epsilon(r) = \left( \frac{m}{8\pi^{3/2} R_1^3} e^{-\frac{z^2}{4 R_1^2}} \right)\, {}_1F_1\left(1, 1, -\frac{\rho^2}{4 R_1^2} \right)\ .
\end{equation}
Now consider the action of the radial Laplacian on a function ${f\Big(-\frac{\rho^2}{4 R_1^2}\Big)}$,
\begin{equation}
    \nabla_\rho^2 f\left(-\frac{\rho^2}{4 R_1^2}\right) = -\frac{\chi f'' + f'}{R_1^2}\ ,
\end{equation}
where ${\chi = -\frac{\rho^2}{4 R_1^2}}$. If ${f(\chi) = {}_1F_1(a, b, \chi)}$, then it satisfies the differential equation
\begin{equation}
    \chi f'' + (b - \chi) f' = a f\ .
\end{equation}
For the case ${a = b = 1}$, this simplifies to
\begin{equation}
    -\frac{\chi f'' + f'}{R_1^2} = -\frac{\chi f' + f}{R_1^2}\ ,
\end{equation}
which, combined with the identity
\begin{equation}
    \chi \frac{\partial}{\partial \chi} \Big({}_1F_1(a, b, \chi)\Big) + a\Big({}_1F_1(a, b, \chi)\Big) = a\Big( {}_1F_1(a + 1, b, \chi)\Big)\ ,
\end{equation}
yields
\begin{equation}
    \nabla_\rho^2 \epsilon(r) = -\left( \frac{m}{8\pi^{3/2} R_1^3} e^{-\frac{z^2}{4 R_1^2}} \right)\, \frac{1}{R_1^2} {}_1F_1\left(2, 1, -\frac{\rho^2}{4 R_1^2} \right)\ .
\end{equation}
Higher-order derivatives can be obtained recursively, thus establishing Eq.~\eqref{eq:EigenValueRelation} by induction.

Substituting this result into Eq.~\eqref{eq:EMTwithRotationFromNR}, the EMT sourced by a Gaussian structure function and arbitrary form factors takes the form
\begin{widetext}
\begin{equation}\label{eq:GeneralGaussianEMT}
\begin{aligned}
    T^{00}(\rho, z) &= \left( \frac{m}{8\pi^{3/2} R_1^3} e^{-\frac{z^2}{4 R_1^2}} \right) \sum_{\ell=0}^{+\infty} \ell!\, F_{2\ell, 1} \left( \frac{a^2}{R_1^2} \right)^\ell {}_1F_1\left(\ell+1, 1, -\frac{\rho^2}{4 R_1^2} \right)\ , \\
    T^{ij}(\rho, z) &= -\left( \frac{m}{8\pi^{3/2} R_2^3} e^{-\frac{z^2}{4 R_2^2}} \right) (s \times \partial)^i (s \times \partial)^j \sum_{\ell=0}^{+\infty} \ell!\, F_{2\ell+2, 2} \left( \frac{a^2}{R_2^2} \right)^\ell {}_1F_1\left(\ell+1, 1, -\frac{\rho^2}{4 R_2^2} \right)\ , \\
    T^{0i}(\rho, z) &= \frac{1}{2} \left( \frac{m}{8\pi^{3/2} R_3^3} e^{-\frac{z^2}{4 R_3^2}} \right) (s \times \partial)^i \sum_{\ell=0}^{+\infty} \ell!\, F_{2\ell+1, 3} \left( \frac{a^2}{R_3^2} \right)^\ell {}_1F_1\left(\ell+1, 1, -\frac{\rho^2}{4 R_3^2} \right)\ .
\end{aligned}
\end{equation}
\end{widetext}

The above expression provides a closed-form representation of a linearized EMT, valid to all orders in the angular momentum expansion, for a rotating source with a Gaussian-like profile. By construction, it induces a gravitational field with multipole moments matching exactly those given in Eq.~\eqref{eq:GravitationalMultipoles}.

While the construction presented here holds for arbitrary multipole moments encoded in the form factors ${F_{\ell,n}}$, we now specialize to the case in which these are chosen to reproduce exactly the multipolar structure of a Kerr BH. This will allow us to study in detail the properties of a specific mimicker sourced by a rotating anisotropic fluid with a Gaussian profile.

\needspace{5\baselineskip}
\section{Kerr case}\label{sec:KerrCase}

As already pointed out, our goal is to apply the general construction of Eq.~\eqref{eq:EMTwithRotationFromNR} to build a physically reasonable linearized source that mimics the Kerr geometry. Starting from the Gaussian-like source introduced in the previous section, and given that we are free to impose any asymptotic structure, constructing a Kerr mimicker simply requires us to impose the Kerr multipolar structure onto Eq.~\eqref{eq:GeneralGaussianEMT}.

As shown in~\cite{Bianchi:2024shc}, the resummed expression of the form factors corresponding to the Kerr multipoles reads
\begin{equation}\label{eq:KerrFF}
\begin{gathered}
    F_{1}(a q_\perp) + (a q_\perp)^2 F_{2}(a q_\perp) = \cos(a q_\perp)\ , \\
    F_{3}(a q_\perp) = \frac{\sin(a q_\perp)}{a q_\perp}\ .
\end{gathered}
\end{equation}
From Eq.~\eqref{eq:KerrFF}, one sees that the mass and stress form factors are redundant. Indeed, in ${d = 3}$ spatial dimensions, stress form factors (and the associated stress multipoles) do not contribute to the gauge-invariant asymptotic structure of the spacetime. Therefore, we are free to choose them arbitrarily without altering the gravitational multipole content. This defines an equivalence class of EMTs all inducing the same asymptotic geometry.

\subsection{Gaussian-smeared Israel source}

Due to the redundancy of the stress form factor in four-dimensional spacetime, we can fix it conveniently to simplify the EMT in Eq.~\eqref{eq:GeneralGaussianEMT}. A particularly interesting case is obtained by setting ${F_2(a q_\perp) = 0}$, which corresponds to ${T^{ij}(x) = 0}$. We refer to the resulting configuration as the Gaussian-smeared Israel source, since in the limit ${R_n \to 0}$, the EMT reduces to the singular distribution identified by Israel~\cite{Israel:1970kp}.

The original Israel EMT describes a disk of radius ${a}$ rotating at superluminal speed and violating energy and causality conditions. It is singular at ${\rho = a}$ and reproduces the ring curvature singularity of the Kerr solution. Moreover, it has been shown that the Israel EMT sources the linearized Kerr geometry~\cite{Israel:1970kp, Balasin:1993kf, Gambino:2024uge}. We thus expect our Gaussian-smeared generalization, controlled by the scales ${R_n}$, to regularize the ring singularity and recover the Israel/Kerr limit as ${R_n \to 0}$.

To this end, we define two master functions
\begin{widetext}
\begin{equation}\label{eq:MasterIsreaelIntegrals}
\begin{aligned}
    \mathcal{M}(\rho, z; R) &= \left( \frac{m}{8\pi^{3/2} R^3} e^{-\frac{z^2}{4 R^2}} \right) \sum_{\ell=0}^{+\infty} (-1)^\ell \frac{\ell!}{(2\ell)!} \left( \frac{a^2}{R^2} \right)^\ell {}_1F_1\left(\ell+1, 1, -\frac{\rho^2}{4 R^2} \right)\ , \\
    \mathcal{J}(\rho, z; R) &= \left( \frac{m}{8\pi^{3/2} R^3} e^{-\frac{z^2}{4 R^2}} \right) \sum_{\ell=0}^{+\infty} (-1)^\ell \frac{\ell!}{(2\ell+1)!} \left( \frac{a^2}{R^2} \right)^\ell {}_1F_1\left(\ell+1, 1, -\frac{\rho^2}{4 R^2} \right)\ ,
\end{aligned}
\end{equation}
\end{widetext}
so that the EMT reads
\begin{equation}\label{eq:EMTphenoImplicit}
\begin{aligned}
    T^{00}(\rho, z) &= \mathcal{M}(\rho, z; R_1)\ , \\
    T^{0i}(\rho, z) &= \frac{1}{2} (s \times \partial)^i\, \mathcal{J}(\rho, z; R_3)\ , \\
    T^{ij}(\rho, z) &= 0\ .
\end{aligned}
\end{equation}

\subsection{Gaussian-smeared Myers–Perry source}

Although we will focus primarily on the Gaussian-smeared Israel source, it is worth briefly mentioning another natural and well-motivated choice for the stress form factors. In higher dimensions, where the stress multipoles do contribute to the gravitational field, the Myers–Perry BH~\cite{Myers:1986un} provides a generalization to arbitrary dimensions of the Kerr solution. In this case, a fully consistent set of form factors reproducing the multipolar structure of Myers–Perry can be written~\cite{Bianchi:2024shc} as
\begin{equation}\label{eq:FFGenericDimension}
\begin{gathered}
    F_2^{(d)}(\zeta) = -\frac{1}{2 \zeta}\, \mathcal{Z}_1^{(d)}(\zeta)\ , \qquad
    F_3^{(d)}(\zeta) = \mathcal{Z}_0^{(d)}(\zeta)\ , \\
    F_1^{(d)}(\zeta) = \zeta^2 F_2^{(d)}(\zeta) + F_3^{(d)}(\zeta)\ ,
\end{gathered}
\end{equation}
where
\begin{equation}
    \mathcal{Z}_n^{(d)}(\zeta) = \frac{\Gamma(d/2)\, \zeta^{-(d-2)/2}}{2^{2-d} (d-1)^{(d-2)/2}} J_{n + (d-2)/2}\left( \frac{d-1}{2} \zeta \right)\ ,
\end{equation}
with ${J_n}$ denoting Bessel functions of the first kind.

In particular, when applied to ${d = 3}$, these form factors yield
\begin{equation}
\begin{aligned}
    F_1(a q_\perp) &= \frac{1}{2} \left( \cos(a q_\perp) + \frac{\sin(a q_\perp)}{a q_\perp} \right)\ , \\
    F_2(a q_\perp) &= \frac{1}{2} \left( \cos(a q_\perp) - \frac{\sin(a q_\perp)}{a q_\perp} \right)\ , \\
    F_3(a q_\perp) &= \frac{\sin(a q_\perp)}{a q_\perp}\ .
\end{aligned}
\end{equation}
This configuration provides then an alternative EMT, effectively representing the ${d=3}$ limit of a Myers–Perry mimicker source in higher dimensions, reproducing once again the asymptotic structure of Kerr BHs, even though a detailed phenomenological analysis of this and other possible EMTs is left for future work.

In summary, by fixing the form factors to match the Kerr multipolar structure and choosing a Gaussian structure function, we have constructed a smooth, analytic EMT that mimics the linearized Kerr spacetime. In the following section, we analyze the physical properties of this source — such as energy conditions, rotation velocity, and causality — across different parameter configurations.

\needspace{5\baselineskip}
\section{Source phenomenology}\label{sec:EMT_KerrPheno}

In this section, we explicitly express Eq.~\eqref{eq:EMTphenoImplicit} in cylindrical coordinates and match the resulting EMT to that of an anisotropic rotating fluid, laying the groundwork for a possible non-perturbative generalization. We then explore different parameter configurations, showing that, at linear order, the Gaussian-smeared Israel source can satisfy both energy and causality conditions.

Let us work in cylindrical coordinates ${\vec{x} = (\rho, \phi, z)}$, with
\begin{equation}
    x = \rho \cos\phi\ , \qquad y = \rho \sin\phi\ ,
\end{equation}
where the flat background metric reads ${\eta_{\mu\nu} = \text{diag}(-1, 1, \rho^2, 1)}$. Since the spin is aligned along the ${z}$-axis, the tensorial structure of the EMT simplifies to
\begin{equation}\label{eq:EMTansatzPL}
T_{I}^{\mu\nu} =
\begin{pNiceMatrix}[columns-width=10pt]
    \mathcal{M}(R_1) & 0 & \frac{a}{2\rho}\partial_\rho \mathcal{J}(R_3) & 0 \\
    0 & 0 & 0 & 0 \\
    \frac{a}{2\rho}\partial_\rho \mathcal{J}(R_3) & 0 & 0 & 0 \\
    0 & 0 & 0 & 0
\end{pNiceMatrix}\ .
\end{equation}
By construction, the source shares the same symmetries as the Kerr geometry, namely axial and equatorial symmetry, so the EMT is independent of ${\phi}$. Moreover, from here on, we will suppress the ${\rho}$ and ${z}$ dependence in notation, keeping only the explicit dependence on ${R_n}$.

To facilitate a non-perturbative extension via the full Einstein equations, we interpret the source as a covariant rotating anisotropic fluid, modeled by the ansatz
\begin{equation}\label{eq:EMTAnsatz}
    T^{\mu\nu} = \epsilon\, u^\mu u^\nu + p_\rho\, l_\rho^{\mu} l_\rho^{\nu} + p_\phi\, l_\phi^\mu l_\phi^\nu\ ,
\end{equation}
where ${u^\mu = \gamma(1, 0, \Omega, 0)}$ is the fluid four-velocity, normalized as ${u^\mu u_\mu = -1}$, with ${\gamma = (1 - \rho^2 \Omega^2)^{-1/2}}$ the Lorentz factor. Here, ${\Omega = \Omega(\rho, z)}$ is the angular velocity, and ${l_\rho^\mu = (0, 1, 0, 0)}$, ${l_\phi^\mu = \gamma(\rho \Omega, 0, 1/\rho, 0)}$ are unit space-like vectors orthogonal to ${u^\mu}$ and to each other, aligned with the ${\rho}$ and ${\phi}$ directions respectively. The eigenvalues of $T^\mu{}_\nu$ are then identified as the energy density ${\epsilon}$ and the anisotropic pressures ${p_\rho}$ and ${p_\phi}$.

We now specify what are the positive energy and causality conditions we are going to consider. The weak energy condition is defined by requiring ${T^{\mu\nu} U_\mu U_\nu \geq 0}$ for any timelike vector ${U^\mu = \alpha_1 u^\mu + \alpha_2 l_\rho^\mu + \alpha_3 l_\phi^\mu}$ satisfying ${U^\mu U_\mu = -1}$, which implies ${\alpha_1^2 = 1 + \alpha_2^2 + \alpha_3^2}$. Substituting into Eq.~\eqref{eq:EMTAnsatz}, we obtain the conditions
\begin{equation}\label{eq:EnergyCondition}
    \epsilon \geq 0\ , \qquad \xi_\rho = \epsilon + p_\rho \geq 0\ , \qquad \xi_\phi = \epsilon + p_\phi \geq 0\ .
\end{equation}
Causality on the other hand, requires all characteristic speeds to be subluminal. The tangential rotational speed is ${v = \rho \Omega}$, and causality imposes ${|v| < 1}$. Additionally, the sound speeds, defined as ${c_k^2 = \partial p_k / \partial \epsilon}$ for ${k = \rho, \phi}$, must also satisfy ${c_k^2 < 1}$. In certain configurations however, we will encounter regions where ${c_k^2 < 0}$, signaling a linear instability of the fluid~\cite{Rezzolla:2013dea, Romatschke:2017ejr}. We will not treat this as a definitive rule-out condition, since stability might be restored in the full non-perturbative regime.

Matching Eq.~\eqref{eq:EMTansatzPL} with the fluid ansatz in Eq.~\eqref{eq:EMTAnsatz}, we obtain the following relations
\begin{equation}\label{eq:FuncDef}
\begin{gathered}
    \epsilon = \frac{\mathcal{M}(R_1) + \sqrt{ \mathcal{M}(R_1)^2 - \left( a \partial_\rho \mathcal{J}(R_3) \right)^2 }}{2}\ , \\
    \Omega = \frac{\mathcal{M}(R_1) - \sqrt{ \mathcal{M}(R_1)^2 - \left( a \partial_\rho \mathcal{J}(R_3) \right)^2 }}{a \rho\, \partial_\rho \mathcal{J}(R_3)}\ , \\
    p_\phi = \frac{ -\mathcal{M}(R_1) + \sqrt{ \mathcal{M}(R_1)^2 - \left( a \partial_\rho \mathcal{J}(R_3) \right)^2 }}{2}\ , \\
    p_\rho = 0\ ,
\end{gathered}
\end{equation}
and for these quantities to be real, the following condition must hold
\begin{equation}\label{eq:RealityCond}
    \mathcal{M}(R_1)^2 \geq \left( a \partial_\rho \mathcal{J}(R_3) \right)^2\ .
\end{equation}
To satisfy this, we parametrize ${R_1 = R}$ and ${R_3 = \alpha R}$, and find that the inequality is satisfied for all ${\rho}$ when ${\alpha < 1}$ and ${R > R^*_\alpha}$ for some threshold ${R^*_\alpha}$. 

\begin{figure}[htbp!]
\centering
\includegraphics[width=0.48\textwidth, valign=c]{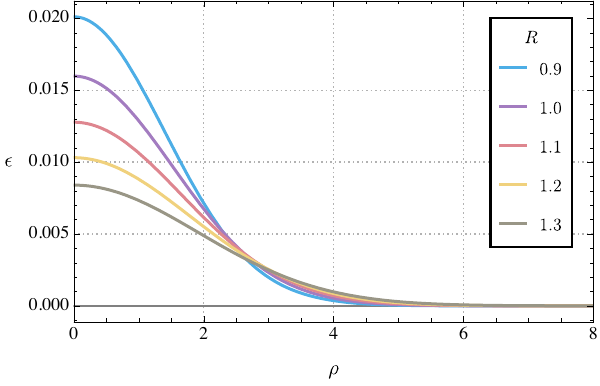}
\includegraphics[width=0.48\textwidth, valign=c]{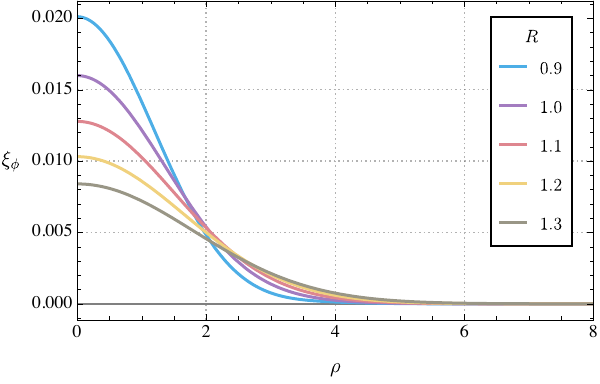}
\caption{On the top the energy density and on the bottom the $\xi_\phi$ variable for the study of the energy condition of the Gaussian-smeared Israel source for ${a=0.8}$, ${z=0}$ and ${\alpha=0.99}$ in units of ${G_N=m=1}$.}
\label{fig:EnergyCondA099}
\end{figure}

The first case of study is for ${\alpha=0.99}$, just below the unity threshold value. Fixing the angular momentum density to a reference value of ${a=0.8}$, the weak energy condition, and hence the energy-density distribution, are studied in Fig.~\ref{fig:EnergyCondA099} for different values of the $R$ parameter. For this configuration the threshold value for which Eq.~\eqref{eq:RealityCond} is satisfied is ${R_{\alpha=0.99}^*\approx 0.84}$, so then we will consider only values ${R>R_{\alpha=0.99}^*}$. Moreover notice that from here on we will study the source phenomenology keeping the angular momentum fixed for different values of $R$ and $\alpha$, since in Eq.~\eqref{eq:MasterIsreaelIntegrals} we can see that the angular momentum always enter in a ratio with the $R$ parameter. Likewise, the dependency on $z$ in Eq.~\eqref{eq:MasterIsreaelIntegrals} enters just as a Gaussian damping factor, hence, for simplicity, we will always consider ${z=0}$ in the study of the EMT. From Fig.~\ref{fig:EnergyCondA099} one can see that the energy condition is satisfied since both $\epsilon$ and $\xi_\phi$ are always positive. Moreover, the energy-density distribution preserves a Gaussian-like profile even in the presence of rotation and increases its central value with $R$ becoming smaller

\begin{figure}[htbp!]
\centering
\includegraphics[width=0.48\textwidth, valign=c]{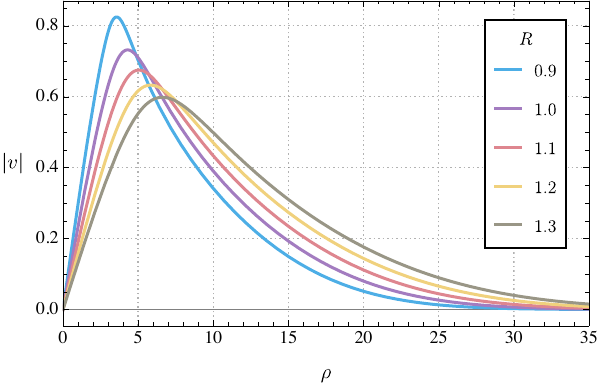}
\includegraphics[width=0.48\textwidth, valign=c]{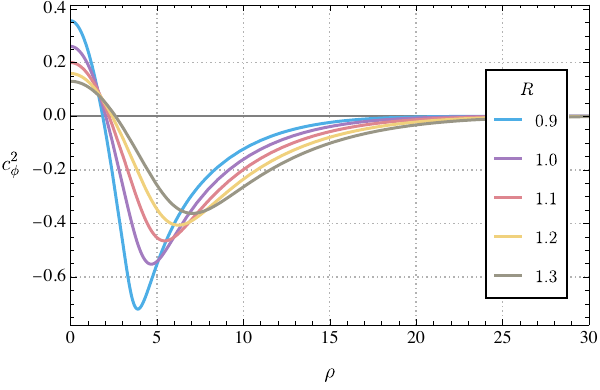}
\caption{On the top the tangential rotational speed and on the bottom the sound speed in the $\phi$-direction of the Gaussian-smeared Israel source for ${a=0.8}$, ${z=0}$ and ${\alpha=0.99}$ in units of ${G_N=m=1}$.}
\label{fig:CausalityCondA099}
\end{figure}

Causality is tested in Fig.~\ref{fig:CausalityCondA099}, where we plot the rotational tangential speed and the sound speed.
The rotational speed satisfies ${|v| < 1}$ throughout and decays to zero at large ${\rho}$. Moreover, as ${R \rightarrow R^*_{\alpha=0.99}}$, the maximum rotational speed approaches unity for some $\rho$ value. On the other hand, the sound speed develops an imaginary part. Indeed, its real part remains subluminal, while the imaginary part develops a peak around $\rho\approx 4R$ and decays at large distances. While this suggests an instability, we do not exclude such configurations, anticipating that non-linear effects could restore stability. 

It is crucial to emphasize that, within the presented model, this instability arises for every parameter choice due to the intrinsic behavior of the energy density and tangential pressure. Specifically, since ${\epsilon}$ monotonically decreases while ${p_\phi}$ vanishes at the origin and exhibits a stationary point at ${r > 0}$, then inevitably exists a critical radius beyond which ${c_\phi^2 < 0}$ holds. Thus, a completely "safe" parameter region without such instabilities is absent within the current linearized description.

\begin{figure}[htbp!]
\centering
\includegraphics[width=0.48\textwidth, valign=c]{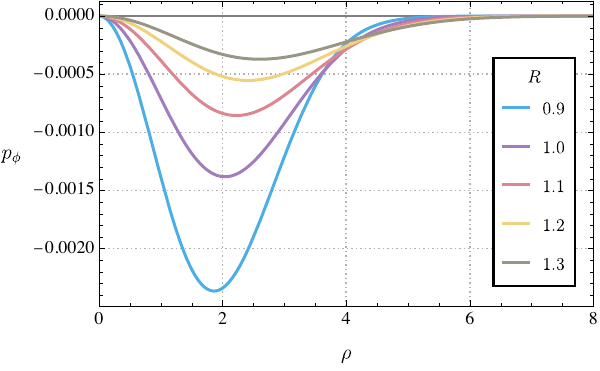}
\caption{Pressure in the $\phi$-direction of the Gaussian-smeared Israel source with ${a = 0.8}$, ${z = 0}$ and ${\alpha = 0.99}$ in units of ${G_N=m = 1}$.}
\label{fig:PressureA099}
\end{figure}

Finally the pressure profile is shown in Fig.~\ref{fig:PressureA099}.
We can see that the pressure vanishes at the origin and reaches a peak around ${\rho \approx 2 R}$, similarly to the rotational velocity phenomenology as expected. Indeed, the behavior of these quantities suggest the source to have a shell-like shape (ring-like in the equatorial plane), typical of regularized Kerr-cores.

\begin{figure*}[h]
\centering
\includegraphics[width=0.48\textwidth, valign=c]{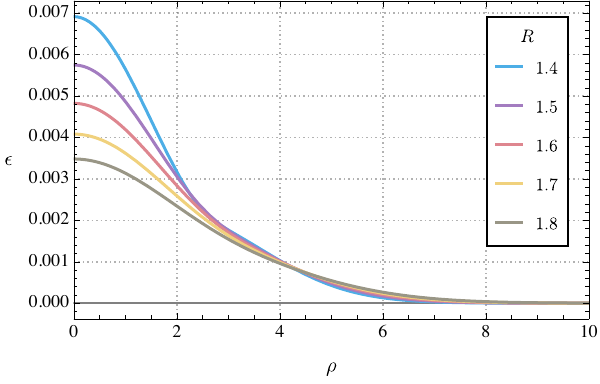}
\includegraphics[width=0.48\textwidth, valign=c]{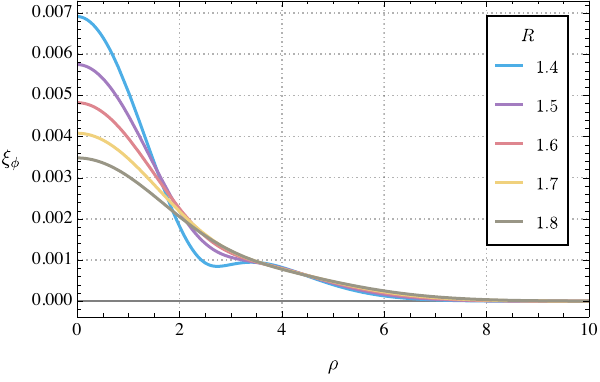}
\includegraphics[width=0.48\textwidth, valign=c]{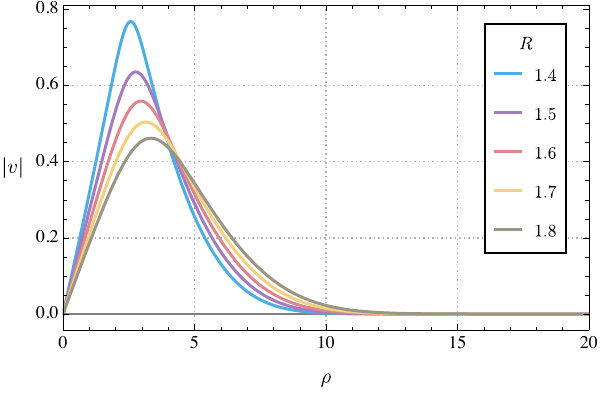}
\includegraphics[width=0.48\textwidth, valign=c]{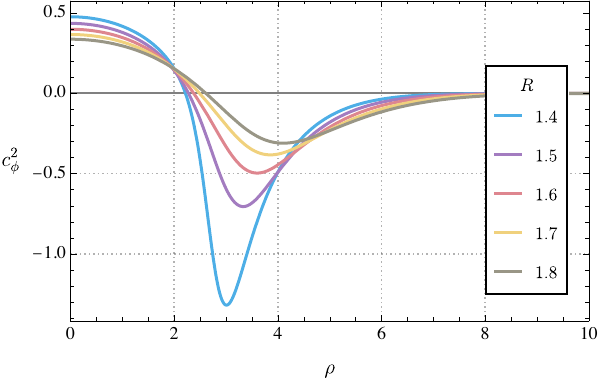}
\includegraphics[width=0.48\textwidth, valign=c]{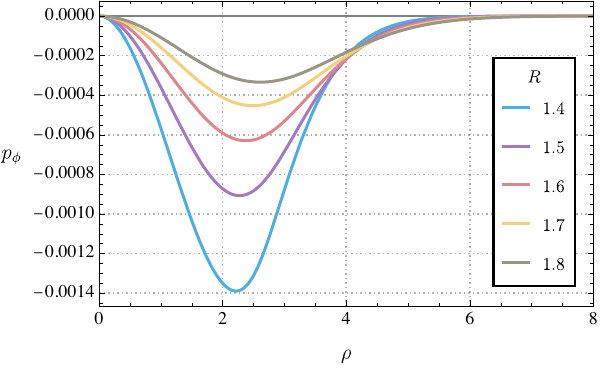}
\caption{On the top-left the energy density, on the top-right the $\xi_\phi$ variable for the study of the energy condition, on the mid-left the tangential rotational speed, on the mid-right the sound speed in the $\phi$-direction and on the bottom the pressure in the $\phi$-direction of the Gaussian-smeared Israel source for ${a=0.8}$, ${z=0}$ and ${\alpha=0.8}$ in units of ${G_N=m=1}$.}
\label{fig:PlotA08}
\end{figure*}

We now consider the case for ${\alpha = 0.8}$, with the corresponding results shown in Fig.~\ref{fig:PlotA08}. In this case, the threshold is ${R^*_{\alpha=0.8} \approx 1.35}$, and since we are only interested in physically viable configurations we consider only the parameter space for which ${R>R_{\alpha=0.8}^*}$.
As in the previous case, all energy and causality conditions are satisfied. Moreover, the larger threshold ${R^*_{\alpha=0.8} > R^*_{\alpha=0.99}}$ reflects the fact that when ${R_1 > R_3}$, their difference must be balanced by larger absolute values to maintain real-valued quantities. Ultimately, we can conclude that the phenomenology remains qualitatively similar, with differences only in peak amplitudes and spatial profiles.

In conclusion, the Gaussian-smeared Israel source defined in Eqs.~\eqref{eq:EMTAnsatz} and~\eqref{eq:FuncDef} provides, within a certain parameter range, a physically viable EMT that induces a gravitational field with the exact Kerr multipolar structure. Although we have focused on a particular example, the general procedure described in Sec.~\ref{sec:GeneralConstruction} is broadly applicable, and other satisfactory mimickers could be constructed by varying the structure functions or the form factor choices. Ultimately, this framework lays the foundation for developing a full non-perturbative solution of Einstein's equations. Indeed, in the static limit (${a = 0}$), an exact solution is already known~\cite{Nicolini:2005vd}, demonstrating that the Gaussian-smeared Israel source admits a global completion at least in the non-rotating case.

\needspace{5\baselineskip}
\section{Metric phenomenology}\label{sec:MetricPheno}

Having established that the Gaussian-smeared Israel source defines a physically reasonable configuration at linear order, it is instructive to explore the phenomenology of the resulting metric. Although the construction is valid only at linearized level, this approximation dominates the long-range expansion of the full spacetime. 

Consider the EMT in Eq.~\eqref{eq:EMTphenoImplicit} defined onto a flat background spacetime, where
\begin{equation}
    T = \eta^{\mu\nu} T_{\mu\nu} = -T^{00}\ .
\end{equation}
Working in cylindrical coordinates and imposing the harmonic gauge, the projector becomes
\begin{equation}
    P_{\mu\nu, \rho\sigma} = \frac{1}{2} \left( \eta_{\mu\rho} \eta_{\nu\sigma} + \eta_{\mu\sigma} \eta_{\nu\rho} - \eta_{\mu\nu} \eta_{\rho\sigma} \right)\ ,
\end{equation}
so that the components of the linearized metric in Eq.~\eqref{eq:FormalMetricfromEMT} take the form
\begin{equation}\label{eq:MetricImplicitComponents}
\begin{aligned}
    h_{00} &= 8\pi G \int \frac{d^3 q}{(2\pi)^3} e^{-i q \cdot x} \frac{1}{q^2} T_{00}(q)\ , \\
    h_{0\phi} &= 8\pi G \int \frac{d^3 q}{(2\pi)^3} e^{-i q \cdot x} \frac{1}{q^2} T_{0\phi}(q)\ , \\
    h_{ij} &= \eta_{ij} h_{00}\ .
\end{aligned}
\end{equation}

As expected, in four spacetime dimensions the stress multipoles vanish, so the linearized metric has only two independent components, namely temporal and angular sectors. We now examine them separately.

\subsection{Temporal component}

Substituting the momentum-space EMT into Eq.~\eqref{eq:MetricImplicitComponents}, the temporal component becomes
\begin{equation}
    h_{00} = 8\pi G m \int \frac{d^3 q}{(2\pi)^3} e^{-i q \cdot x} e^{-q^2 R^2} \cos(a q_\perp)\ .
\end{equation}
Switching to cylindrical coordinates and integrating over the angular part, we obtain
\begin{widetext}
\begin{equation}
    h_{00} = 8\pi G m \int_{-\infty}^{+\infty} \frac{d q_z}{2\pi} e^{-i q_z z} \int_0^{+\infty} \frac{d q_\perp\, q_\perp}{2\pi} J_0(q_\perp \rho) \frac{e^{-q_z^2 R^2}}{q_z^2 + q_\perp^2} e^{-q_\perp^2 R^2} \cos(a q_\perp)\ .
\end{equation}
The ${q_z}$ integral can be performed analytically, resulting in a single-integral expression for the temporal component as in
\begin{equation}\label{eq:TemporalMetricMimicker}
    h_{00} = G m \int d q_\perp\, J_0(q_\perp \rho)\, \cos(a q_\perp)\, \left[ e^{-q_\perp z} \mathrm{Erfc}\left(q_\perp R - \tfrac{z}{2R}\right) + e^{q_\perp z} \mathrm{Erfc}\left(q_\perp R + \tfrac{z}{2R} \right) \right]\ ,
\end{equation}
\end{widetext}
where ${\mathrm{Erfc}(x)}$ is the complementary error function
\begin{equation}
    \mathrm{Erfc}(x) = 1 - \frac{2}{\pi} \int_0^x dt\, e^{-t^2}\ .
\end{equation}

As far as we know, the integral in Eq.~\eqref{eq:TemporalMetricMimicker} does not admit a closed-form expression, and must be evaluated numerically. However, in the ${R \to 0}$ limit, corresponding to the Kerr singular case, the integral simplifies to
\begin{equation}\label{eq:KerrTempImp}
    h_{00}^{Kerr} = 2G m \int d q_\perp\, J_0(q_\perp \rho)\, \cos(a q_\perp)\, e^{-q_\perp |z|}\ ,
\end{equation}
which can be analytically integrated by means of tabulated integrals listed in~\cite{Gradshteyn:1943cpj}, leading to  
\begin{equation}\label{eq:MetricMasterInt}
   h_{00}^{Kerr}=2Gm\frac{\sqrt{l_+^2-a^2}}{l_+^2-l_-^2}\ ,
\end{equation}
where
\begin{equation}\label{eq:lsDef}
\begin{aligned}
    l_+&=\frac{\sqrt{(a+\rho)^2+z^2}+\sqrt{(a-\rho)^2+z^2}}{2}\ ,\\ 
    l_-&=\frac{\sqrt{(a+\rho)^2+z^2}-\sqrt{(a-\rho)^2+z^2}}{2}\ .
\end{aligned}
\end{equation}

\begin{figure*}[htbp!]
\centering
\includegraphics[width=0.48\textwidth, valign=c]{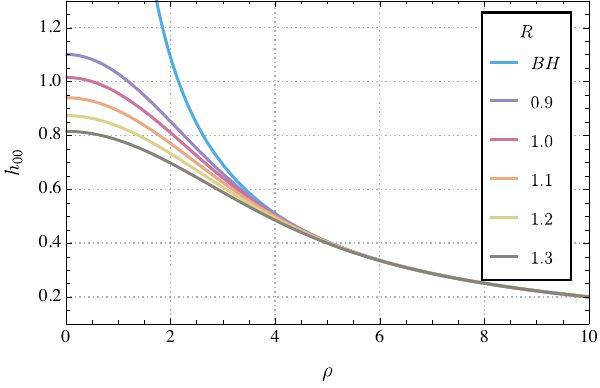}
\includegraphics[width=0.48\textwidth, valign=c]{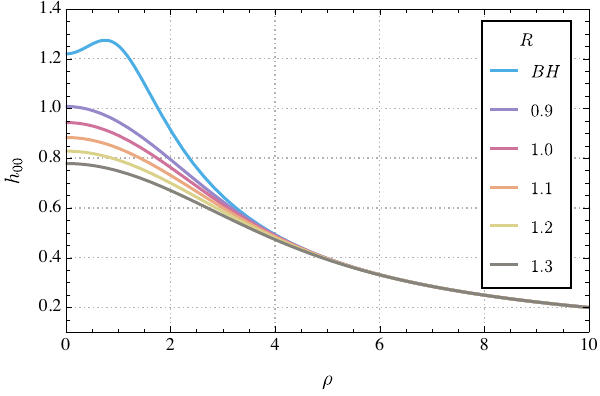}
\caption{Temporal component of the linearized metric sourced by the Gaussian-smeared Israel EMT for different values of ${R}$ and for ${z = 0}$ on the left and ${z = 1}$ on the right, with ${a = 0.8}$ and in units of ${G_N=m=1}$.}
\label{fig:Metric00}
\end{figure*}

This metric component is singular on the ring ${z = 0}$ and ${\rho = a}$, corresponding to the Kerr curvature singularity. For finite ${R}$ however, this singularity is smeared and a comparison between the exact Kerr and the mimicking metric is shown in Fig.~\ref{fig:Metric00}.
As expected, the mimicking metric approaches the Kerr profile asymptotically, and for finite ${R}$, the curvature singularity is smeared into a smooth core. We can see that the central value of the metric depends on the $R$ parameter, and that differently from the EMT study the dependence on $z$ is no more a simple dumping factor, hence it is worth depicting the gravitational field for different slices of $z$.

\subsection{Angular component}

We now consider the angular component of the metric, which in integral form is given by
\begin{widetext}
\begin{equation}
    h_{0\phi} = -4\pi G m a \rho\, \partial_\rho \int \frac{d q_z}{2\pi} e^{-i q_z z} \int \frac{d q_\perp\, q_\perp}{2\pi} J_0(q_\perp \rho)\, \frac{e^{-q_z^2 R^2}}{q_z^2 + q_\perp^2}\, e^{-q_\perp^2 R^2} \frac{\sin(a q_\perp)}{a q_\perp}\ .
\end{equation}
Again, the ${q_z}$ integration can be carried out analytically, yielding
\begin{equation}\label{eq:AngularPart}
    h_{0\phi} = \frac{G m \rho}{2} \int d q_\perp\, J_1(q_\perp \rho)\, \sin(a q_\perp)\, \left[ e^{-q_\perp z} \mathrm{Erfc}\left(q_\perp R - \tfrac{z}{2R} \right) + e^{q_\perp z} \mathrm{Erfc}\left(q_\perp R + \tfrac{z}{2R} \right) \right]\ .
\end{equation}
\end{widetext}

As with the temporal component, no closed-form solution is known for Eq.~\eqref{eq:AngularPart}, however, in the ${R \to 0}$ limit the angular component simplifies to
\begin{equation}
    h_{0\phi}^{Kerr} = G m \rho \int d q_\perp\, J_1(q_\perp \rho)\, \sin(a q_\perp)\, e^{-q_\perp |z|}\ ,
\end{equation}
which can be integrated analytically by using known tabulated integrals~\cite{Gradshteyn:1943cpj} as in 
    \begin{equation}
        h_{0\phi}^{Kerr} =  \frac{G m \rho^2 a}{l_+^2}\frac{\sqrt{l_+^2-a^2}}{l_+^2-l_-^2}\ ,
    \end{equation}
where $l_+$ and $l_-$ are defined in Eq.~\eqref{eq:lsDef}.

\begin{figure*}[htbp!]
\centering
\includegraphics[width=0.48\textwidth, valign=c]{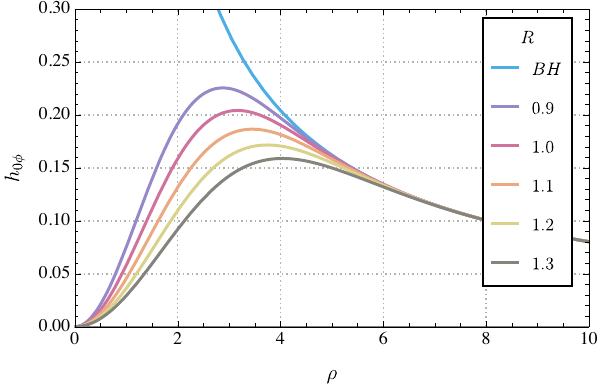}
\includegraphics[width=0.48\textwidth, valign=c]{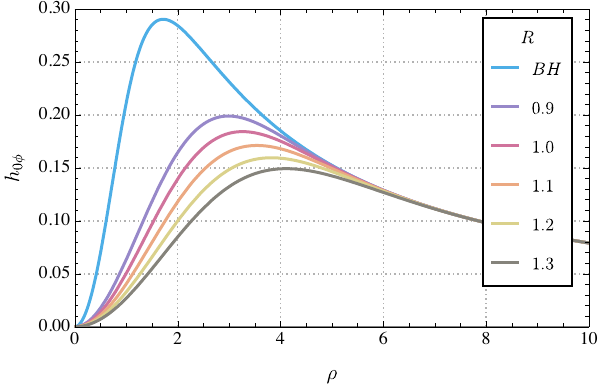}
\caption{Angular component of the linearized metric sourced by the Gaussian-smeared Israel EMT for different values of ${\alpha R}$ and for ${z = 0}$ on the left and ${z = 1}$ on the right, with ${a = 0.8}$ and in units of ${G_N=m=1}$.}
\label{fig:Metric0phi}
\end{figure*}

The behavior of the angular component for finite ${R}$ is then shown in Fig.~\ref{fig:Metric0phi}.
As in the temporal sector, the mimicking metric converges to the Kerr solution asymptotically, and for finite ${R}$, the ring singularity is regularized into a smooth, extended geometry. We can notice that the central value of the angular metric component is always vanishing, while the peak depends parametrically on $R$.

\needspace{5\baselineskip}
\section{Conclusions}\label{sec:Conclusions}

Using the momentum-space formalism of GR, we have developed a general framework for constructing non-singular EMTs that source gravitational fields with any prescribed multipolar structure, culminating with Eq.~\eqref{eq:EMTwithRotationFromNR} that parametrizes multipoles with form factors $F_{\ell, n}$ and the internal structure with structure functions $K_n$. Within this framework, we built a linearized EMT with a Gaussian-like energy-density profile describing an anisotropic rotating fluid that reproduces the exact multipolar structure of a Kerr BH. This method is fully general~\cite{Bianchi:2024shc}, it extends to higher dimensions, and allows for the construction of infinitely many physically distinct sources that share the same asymptotic spacetime — thus defining an equivalence class of EMTs.

Among these, we focused on the Gaussian-smeared Israel source, which provided concrete insight into the viability of the framework. We demonstrated that, at linear order, this configuration constitutes a physically reasonable source for a Kerr mimicker. The source shares the symmetries of the Kerr geometry and is characterized by its mass, angular momentum, and two additional length scales, parametrized as ${R_1 = R}$ and ${R_3 = \alpha R}$. For ${\alpha \geq 1}$, the EMT becomes complex-valued in some regions and is therefore unphysical. However, for ${\alpha < 1}$, we identified regions of parameter space in which the EMT remains real-valued everywhere and is thus physically admissible. Specifically, we found ${R > R^*_{\alpha=0.99} \approx 0.84}$ and ${R > R^*_{\alpha=0.8} \approx 1.35}$ for the two benchmark cases we analyzed.

Within these bounds, both configurations satisfy the weak energy condition and causality. Although the sound speed becomes imaginary in some regions — a potential sign of instability under linearized pressure perturbations~\cite{Rezzolla:2013dea, Romatschke:2017ejr} — we did not interpret this as a definitive exclusion. Indeed, while the argument presented here is valid at all orders in angular momentum, it remains perturbative in $G_N$, and a full non-perturbative generalization may stabilize the source via nonlinear gravitational effects. We also analyzed the linearized metric induced by the mimicking EMT and showed explicitly how the curvature singularity of Kerr is smeared at finite values of the ${R}$ parameter.

It is important to emphasize that the model studied here is only one specific realization within a much broader framework. Different choices for the stress form factor and/or the structure functions may lead to alternative mimickers with distinct phenomenologies. A systematic study of the full equivalence class of Kerr-mimicking EMTs could uncover direct relations between the viability of a source and the analytic structure of its defining functions — potentially enabling a faster and more efficient construction of BH mimickers.

Although a full non-perturbative extension of the Gaussian-smeared Israel source remains a challenging task, one promising approach would be to work order by order in angular momentum, building a nonlinear solution through a multipolar expansion. Indeed, the static case with a Gaussian profile is already known to admit a fully consistent non-perturbative solution~\cite{Nicolini:2005vd}, which supports the plausibility of extending the model to rotation.

Finally, future work could explore whether, and under what conditions, these mimickers develop horizons. The model presented here in fact may offer valuable insights into the physics of regular BHs and the transition from horizonless compact objects to regular BH geometries. In particular, the existence of a lower bound on ${R}$ for physical viability hints at a natural mechanism through which mimickers might evolve into regular BHs — potentially triggered by the breakdown of energy or causality conditions. Understanding this threshold could shed light on how gravitational collapse proceeds in scenarios beyond classical singularities, and on the true nature of compact objects in quantum gravity.

\begin{acknowledgments}

I would like to thank Massimo Bianchi, Paolo Pani, Fabio Riccioni and Roberto Emparan for their useful comments and suggestions throughout the writing process of this paper, and Simone D'Onofrio for insightful discussions at the time this project started. 
I also acknowledge the hospitality of the Universitat de Barcelona/ICCUB and IFAE(Institut de Física d'Altes Energies) during the final stages of this paper's completion.
This work is partially supported by Sapienza University of Rome (``Progetti per Avvio alla Ricerca - Tipo 1'',
protocol number AR1241906DC8FF32).

\end{acknowledgments}

\appendix

\bibliography{biblio}

\begin{thebibliography}{64}%
\makeatletter
\providecommand \@ifxundefined [1]{%
 \@ifx{#1\undefined}
}%
\providecommand \@ifnum [1]{%
 \ifnum #1\expandafter \@firstoftwo
 \else \expandafter \@secondoftwo
 \fi
}%
\providecommand \@ifx [1]{%
 \ifx #1\expandafter \@firstoftwo
 \else \expandafter \@secondoftwo
 \fi
}%
\providecommand \natexlab [1]{#1}%
\providecommand \enquote  [1]{``#1''}%
\providecommand \bibnamefont  [1]{#1}%
\providecommand \bibfnamefont [1]{#1}%
\providecommand \citenamefont [1]{#1}%
\providecommand \href@noop [0]{\@secondoftwo}%
\providecommand \href [0]{\begingroup \@sanitize@url \@href}%
\providecommand \@href[1]{\@@startlink{#1}\@@href}%
\providecommand \@@href[1]{\endgroup#1\@@endlink}%
\providecommand \@sanitize@url [0]{\catcode `\\12\catcode `\$12\catcode `\&12\catcode `\#12\catcode `\^12\catcode `\_12\catcode `\%12\relax}%
\providecommand \@@startlink[1]{}%
\providecommand \@@endlink[0]{}%
\providecommand \url  [0]{\begingroup\@sanitize@url \@url }%
\providecommand \@url [1]{\endgroup\@href {#1}{\urlprefix }}%
\providecommand \urlprefix  [0]{URL }%
\providecommand \Eprint [0]{\href }%
\providecommand \doibase [0]{http://dx.doi.org/}%
\providecommand \selectlanguage [0]{\@gobble}%
\providecommand \bibinfo  [0]{\@secondoftwo}%
\providecommand \bibfield  [0]{\@secondoftwo}%
\providecommand \translation [1]{[#1]}%
\providecommand \BibitemOpen [0]{}%
\providecommand \bibitemStop [0]{}%
\providecommand \bibitemNoStop [0]{.\EOS\space}%
\providecommand \EOS [0]{\spacefactor3000\relax}%
\providecommand \BibitemShut  [1]{\csname bibitem#1\endcsname}%
\let\auto@bib@innerbib\@empty
\bibitem [{\citenamefont {Schwarzschild}(1916)}]{Schwarzschild:1916uq}%
  \BibitemOpen
  \bibfield  {author} {\bibinfo {author} {\bibfnamefont {K.}~\bibnamefont {Schwarzschild}},\ }\href@noop {} {\bibfield  {journal} {\bibinfo  {journal} {Sitzungsber. Preuss. Akad. Wiss. Berlin (Math. Phys. )}\ }\textbf {\bibinfo {volume} {1916}},\ \bibinfo {pages} {189} (\bibinfo {year} {1916})},\ \Eprint {http://arxiv.org/abs/physics/9905030} {arXiv:physics/9905030} \BibitemShut {NoStop}%
\bibitem [{\citenamefont {Kerr}(1963)}]{Kerr:1963ud}%
  \BibitemOpen
  \bibfield  {author} {\bibinfo {author} {\bibfnamefont {R.~P.}\ \bibnamefont {Kerr}},\ }\href {\doibase 10.1103/PhysRevLett.11.237} {\bibfield  {journal} {\bibinfo  {journal} {Phys. Rev. Lett.}\ }\textbf {\bibinfo {volume} {11}},\ \bibinfo {pages} {237} (\bibinfo {year} {1963})}\BibitemShut {NoStop}%
\bibitem [{\citenamefont {Misner}\ \emph {et~al.}(1973)\citenamefont {Misner}, \citenamefont {Thorne},\ and\ \citenamefont {Wheeler}}]{Misner:1973prb}%
  \BibitemOpen
  \bibfield  {author} {\bibinfo {author} {\bibfnamefont {C.~W.}\ \bibnamefont {Misner}}, \bibinfo {author} {\bibfnamefont {K.~S.}\ \bibnamefont {Thorne}}, \ and\ \bibinfo {author} {\bibfnamefont {J.~A.}\ \bibnamefont {Wheeler}},\ }\href@noop {} {\emph {\bibinfo {title} {{Gravitation}}}}\ (\bibinfo  {publisher} {W. H. Freeman},\ \bibinfo {address} {San Francisco},\ \bibinfo {year} {1973})\BibitemShut {NoStop}%
\bibitem [{\citenamefont {Abbott}\ \emph {et~al.}(2016)\citenamefont {Abbott} \emph {et~al.}}]{LIGOScientific:2016aoc}%
  \BibitemOpen
  \bibfield  {author} {\bibinfo {author} {\bibfnamefont {B.~P.}\ \bibnamefont {Abbott}} \emph {et~al.} (\bibinfo {collaboration} {LIGO Scientific, Virgo}),\ }\href {\doibase 10.1103/PhysRevLett.116.061102} {\bibfield  {journal} {\bibinfo  {journal} {Phys. Rev. Lett.}\ }\textbf {\bibinfo {volume} {116}},\ \bibinfo {pages} {061102} (\bibinfo {year} {2016})},\ \Eprint {http://arxiv.org/abs/1602.03837} {arXiv:1602.03837 [gr-qc]} \BibitemShut {NoStop}%
\bibitem [{\citenamefont {Abbott}\ \emph {et~al.}(2019)\citenamefont {Abbott} \emph {et~al.}}]{LIGOScientific:2019fpa}%
  \BibitemOpen
  \bibfield  {author} {\bibinfo {author} {\bibfnamefont {B.~P.}\ \bibnamefont {Abbott}} \emph {et~al.} (\bibinfo {collaboration} {LIGO Scientific, Virgo}),\ }\href {\doibase 10.1103/PhysRevD.100.104036} {\bibfield  {journal} {\bibinfo  {journal} {Phys. Rev. D}\ }\textbf {\bibinfo {volume} {100}},\ \bibinfo {pages} {104036} (\bibinfo {year} {2019})},\ \Eprint {http://arxiv.org/abs/1903.04467} {arXiv:1903.04467 [gr-qc]} \BibitemShut {NoStop}%
\bibitem [{\citenamefont {Akiyama}\ \emph {et~al.}(2019)\citenamefont {Akiyama} \emph {et~al.}}]{EventHorizonTelescope:2019dse}%
  \BibitemOpen
  \bibfield  {author} {\bibinfo {author} {\bibfnamefont {K.}~\bibnamefont {Akiyama}} \emph {et~al.} (\bibinfo {collaboration} {Event Horizon Telescope}),\ }\href {\doibase 10.3847/2041-8213/ab0ec7} {\bibfield  {journal} {\bibinfo  {journal} {Astrophys. J. Lett.}\ }\textbf {\bibinfo {volume} {875}},\ \bibinfo {pages} {L1} (\bibinfo {year} {2019})},\ \Eprint {http://arxiv.org/abs/1906.11238} {arXiv:1906.11238 [astro-ph.GA]} \BibitemShut {NoStop}%
\bibitem [{\citenamefont {Akiyama}\ \emph {et~al.}(2022)\citenamefont {Akiyama} \emph {et~al.}}]{EventHorizonTelescope:2022xqj}%
  \BibitemOpen
  \bibfield  {author} {\bibinfo {author} {\bibfnamefont {K.}~\bibnamefont {Akiyama}} \emph {et~al.} (\bibinfo {collaboration} {Event Horizon Telescope}),\ }\href {\doibase 10.3847/2041-8213/ac6756} {\bibfield  {journal} {\bibinfo  {journal} {Astrophys. J. Lett.}\ }\textbf {\bibinfo {volume} {930}},\ \bibinfo {pages} {L17} (\bibinfo {year} {2022})},\ \Eprint {http://arxiv.org/abs/2311.09484} {arXiv:2311.09484 [astro-ph.HE]} \BibitemShut {NoStop}%
\bibitem [{\citenamefont {Bardeen}\ \emph {et~al.}(1973)\citenamefont {Bardeen}, \citenamefont {Carter},\ and\ \citenamefont {Hawking}}]{Bardeen:1973gs}%
  \BibitemOpen
  \bibfield  {author} {\bibinfo {author} {\bibfnamefont {J.~M.}\ \bibnamefont {Bardeen}}, \bibinfo {author} {\bibfnamefont {B.}~\bibnamefont {Carter}}, \ and\ \bibinfo {author} {\bibfnamefont {S.~W.}\ \bibnamefont {Hawking}},\ }\href {\doibase 10.1007/BF01645742} {\bibfield  {journal} {\bibinfo  {journal} {Commun. Math. Phys.}\ }\textbf {\bibinfo {volume} {31}},\ \bibinfo {pages} {161} (\bibinfo {year} {1973})}\BibitemShut {NoStop}%
\bibitem [{\citenamefont {Hawking}(1974)}]{Hawking:1974rv}%
  \BibitemOpen
  \bibfield  {author} {\bibinfo {author} {\bibfnamefont {S.~W.}\ \bibnamefont {Hawking}},\ }\href {\doibase 10.1038/248030a0} {\bibfield  {journal} {\bibinfo  {journal} {Nature}\ }\textbf {\bibinfo {volume} {248}},\ \bibinfo {pages} {30} (\bibinfo {year} {1974})}\BibitemShut {NoStop}%
\bibitem [{\citenamefont {Israel}(1967)}]{Israel:1967wq}%
  \BibitemOpen
  \bibfield  {author} {\bibinfo {author} {\bibfnamefont {W.}~\bibnamefont {Israel}},\ }\href {\doibase 10.1103/PhysRev.164.1776} {\bibfield  {journal} {\bibinfo  {journal} {Phys. Rev.}\ }\textbf {\bibinfo {volume} {164}},\ \bibinfo {pages} {1776} (\bibinfo {year} {1967})}\BibitemShut {NoStop}%
\bibitem [{\citenamefont {Carter}(1971)}]{Carter:1971zc}%
  \BibitemOpen
  \bibfield  {author} {\bibinfo {author} {\bibfnamefont {B.}~\bibnamefont {Carter}},\ }\href {\doibase 10.1103/PhysRevLett.26.331} {\bibfield  {journal} {\bibinfo  {journal} {Phys. Rev. Lett.}\ }\textbf {\bibinfo {volume} {26}},\ \bibinfo {pages} {331} (\bibinfo {year} {1971})}\BibitemShut {NoStop}%
\bibitem [{\citenamefont {Giddings}(1992)}]{Giddings:1992hh}%
  \BibitemOpen
  \bibfield  {author} {\bibinfo {author} {\bibfnamefont {S.~B.}\ \bibnamefont {Giddings}},\ }\href {\doibase 10.1103/PhysRevD.46.1347} {\bibfield  {journal} {\bibinfo  {journal} {Phys. Rev. D}\ }\textbf {\bibinfo {volume} {46}},\ \bibinfo {pages} {1347} (\bibinfo {year} {1992})},\ \Eprint {http://arxiv.org/abs/hep-th/9203059} {arXiv:hep-th/9203059} \BibitemShut {NoStop}%
\bibitem [{\citenamefont {Lunin}\ and\ \citenamefont {Mathur}(2002)}]{Lunin:2001jy}%
  \BibitemOpen
  \bibfield  {author} {\bibinfo {author} {\bibfnamefont {O.}~\bibnamefont {Lunin}}\ and\ \bibinfo {author} {\bibfnamefont {S.~D.}\ \bibnamefont {Mathur}},\ }\href {\doibase 10.1016/S0550-3213(01)00620-4} {\bibfield  {journal} {\bibinfo  {journal} {Nucl. Phys. B}\ }\textbf {\bibinfo {volume} {623}},\ \bibinfo {pages} {342} (\bibinfo {year} {2002})},\ \Eprint {http://arxiv.org/abs/hep-th/0109154} {arXiv:hep-th/0109154} \BibitemShut {NoStop}%
\bibitem [{\citenamefont {Mathur}(2005)}]{Mathur:2005zp}%
  \BibitemOpen
  \bibfield  {author} {\bibinfo {author} {\bibfnamefont {S.~D.}\ \bibnamefont {Mathur}},\ }\href {\doibase 10.1002/prop.200410203} {\bibfield  {journal} {\bibinfo  {journal} {Fortsch. Phys.}\ }\textbf {\bibinfo {volume} {53}},\ \bibinfo {pages} {793} (\bibinfo {year} {2005})},\ \Eprint {http://arxiv.org/abs/hep-th/0502050} {arXiv:hep-th/0502050} \BibitemShut {NoStop}%
\bibitem [{\citenamefont {Hayward}(2006)}]{Hayward:2005gi}%
  \BibitemOpen
  \bibfield  {author} {\bibinfo {author} {\bibfnamefont {S.~A.}\ \bibnamefont {Hayward}},\ }\href {\doibase 10.1103/PhysRevLett.96.031103} {\bibfield  {journal} {\bibinfo  {journal} {Phys. Rev. Lett.}\ }\textbf {\bibinfo {volume} {96}},\ \bibinfo {pages} {031103} (\bibinfo {year} {2006})},\ \Eprint {http://arxiv.org/abs/gr-qc/0506126} {arXiv:gr-qc/0506126} \BibitemShut {NoStop}%
\bibitem [{\citenamefont {Bena}\ and\ \citenamefont {Warner}(2006)}]{Bena:2005va}%
  \BibitemOpen
  \bibfield  {author} {\bibinfo {author} {\bibfnamefont {I.}~\bibnamefont {Bena}}\ and\ \bibinfo {author} {\bibfnamefont {N.~P.}\ \bibnamefont {Warner}},\ }\href {\doibase 10.1103/PhysRevD.74.066001} {\bibfield  {journal} {\bibinfo  {journal} {Phys. Rev. D}\ }\textbf {\bibinfo {volume} {74}},\ \bibinfo {pages} {066001} (\bibinfo {year} {2006})},\ \Eprint {http://arxiv.org/abs/hep-th/0505166} {arXiv:hep-th/0505166} \BibitemShut {NoStop}%
\bibitem [{\citenamefont {Bena}\ \emph {et~al.}(2006)\citenamefont {Bena}, \citenamefont {Wang},\ and\ \citenamefont {Warner}}]{Bena:2006kb}%
  \BibitemOpen
  \bibfield  {author} {\bibinfo {author} {\bibfnamefont {I.}~\bibnamefont {Bena}}, \bibinfo {author} {\bibfnamefont {C.-W.}\ \bibnamefont {Wang}}, \ and\ \bibinfo {author} {\bibfnamefont {N.~P.}\ \bibnamefont {Warner}},\ }\href {\doibase 10.1088/1126-6708/2006/11/042} {\bibfield  {journal} {\bibinfo  {journal} {JHEP}\ }\textbf {\bibinfo {volume} {11}},\ \bibinfo {pages} {042} (\bibinfo {year} {2006})},\ \Eprint {http://arxiv.org/abs/hep-th/0608217} {arXiv:hep-th/0608217} \BibitemShut {NoStop}%
\bibitem [{\citenamefont {Mathur}(2008)}]{Mathur:2008nj}%
  \BibitemOpen
  \bibfield  {author} {\bibinfo {author} {\bibfnamefont {S.~D.}\ \bibnamefont {Mathur}},\ }\href@noop {} {\  (\bibinfo {year} {2008})},\ \Eprint {http://arxiv.org/abs/0810.4525} {arXiv:0810.4525 [hep-th]} \BibitemShut {NoStop}%
\bibitem [{\citenamefont {Frolov}(2016)}]{Frolov:2016pav}%
  \BibitemOpen
  \bibfield  {author} {\bibinfo {author} {\bibfnamefont {V.~P.}\ \bibnamefont {Frolov}},\ }\href {\doibase 10.1103/PhysRevD.94.104056} {\bibfield  {journal} {\bibinfo  {journal} {Phys. Rev. D}\ }\textbf {\bibinfo {volume} {94}},\ \bibinfo {pages} {104056} (\bibinfo {year} {2016})},\ \Eprint {http://arxiv.org/abs/1609.01758} {arXiv:1609.01758 [gr-qc]} \BibitemShut {NoStop}%
\bibitem [{\citenamefont {Cano}\ \emph {et~al.}(2019)\citenamefont {Cano}, \citenamefont {Chimento}, \citenamefont {Ort\'\i{}n},\ and\ \citenamefont {Ruip\'erez}}]{Cano:2018aod}%
  \BibitemOpen
  \bibfield  {author} {\bibinfo {author} {\bibfnamefont {P.~A.}\ \bibnamefont {Cano}}, \bibinfo {author} {\bibfnamefont {S.}~\bibnamefont {Chimento}}, \bibinfo {author} {\bibfnamefont {T.}~\bibnamefont {Ort\'\i{}n}}, \ and\ \bibinfo {author} {\bibfnamefont {A.}~\bibnamefont {Ruip\'erez}},\ }\href {\doibase 10.1103/PhysRevD.99.046014} {\bibfield  {journal} {\bibinfo  {journal} {Phys. Rev. D}\ }\textbf {\bibinfo {volume} {99}},\ \bibinfo {pages} {046014} (\bibinfo {year} {2019})},\ \Eprint {http://arxiv.org/abs/1806.08377} {arXiv:1806.08377 [hep-th]} \BibitemShut {NoStop}%
\bibitem [{\citenamefont {Carballo-Rubio}\ \emph {et~al.}(2018{\natexlab{a}})\citenamefont {Carballo-Rubio}, \citenamefont {Di~Filippo}, \citenamefont {Liberati}, \citenamefont {Pacilio},\ and\ \citenamefont {Visser}}]{Carballo-Rubio:2018pmi}%
  \BibitemOpen
  \bibfield  {author} {\bibinfo {author} {\bibfnamefont {R.}~\bibnamefont {Carballo-Rubio}}, \bibinfo {author} {\bibfnamefont {F.}~\bibnamefont {Di~Filippo}}, \bibinfo {author} {\bibfnamefont {S.}~\bibnamefont {Liberati}}, \bibinfo {author} {\bibfnamefont {C.}~\bibnamefont {Pacilio}}, \ and\ \bibinfo {author} {\bibfnamefont {M.}~\bibnamefont {Visser}},\ }\href {\doibase 10.1007/JHEP07(2018)023} {\bibfield  {journal} {\bibinfo  {journal} {JHEP}\ }\textbf {\bibinfo {volume} {07}},\ \bibinfo {pages} {023} (\bibinfo {year} {2018}{\natexlab{a}})},\ \Eprint {http://arxiv.org/abs/1805.02675} {arXiv:1805.02675 [gr-qc]} \BibitemShut {NoStop}%
\bibitem [{\citenamefont {Simpson}\ and\ \citenamefont {Visser}(2019)}]{Simpson:2018tsi}%
  \BibitemOpen
  \bibfield  {author} {\bibinfo {author} {\bibfnamefont {A.}~\bibnamefont {Simpson}}\ and\ \bibinfo {author} {\bibfnamefont {M.}~\bibnamefont {Visser}},\ }\href {\doibase 10.1088/1475-7516/2019/02/042} {\bibfield  {journal} {\bibinfo  {journal} {JCAP}\ }\textbf {\bibinfo {volume} {02}},\ \bibinfo {pages} {042} (\bibinfo {year} {2019})},\ \Eprint {http://arxiv.org/abs/1812.07114} {arXiv:1812.07114 [gr-qc]} \BibitemShut {NoStop}%
\bibitem [{\citenamefont {Simpson}\ \emph {et~al.}(2019)\citenamefont {Simpson}, \citenamefont {Martin-Moruno},\ and\ \citenamefont {Visser}}]{Simpson:2019cer}%
  \BibitemOpen
  \bibfield  {author} {\bibinfo {author} {\bibfnamefont {A.}~\bibnamefont {Simpson}}, \bibinfo {author} {\bibfnamefont {P.}~\bibnamefont {Martin-Moruno}}, \ and\ \bibinfo {author} {\bibfnamefont {M.}~\bibnamefont {Visser}},\ }\href {\doibase 10.1088/1361-6382/ab28a5} {\bibfield  {journal} {\bibinfo  {journal} {Class. Quant. Grav.}\ }\textbf {\bibinfo {volume} {36}},\ \bibinfo {pages} {145007} (\bibinfo {year} {2019})},\ \Eprint {http://arxiv.org/abs/1902.04232} {arXiv:1902.04232 [gr-qc]} \BibitemShut {NoStop}%
\bibitem [{\citenamefont {Bianchi}\ \emph {et~al.}(2020)\citenamefont {Bianchi}, \citenamefont {Consoli}, \citenamefont {Grillo}, \citenamefont {Morales}, \citenamefont {Pani},\ and\ \citenamefont {Raposo}}]{Bianchi:2020bxa}%
  \BibitemOpen
  \bibfield  {author} {\bibinfo {author} {\bibfnamefont {M.}~\bibnamefont {Bianchi}}, \bibinfo {author} {\bibfnamefont {D.}~\bibnamefont {Consoli}}, \bibinfo {author} {\bibfnamefont {A.}~\bibnamefont {Grillo}}, \bibinfo {author} {\bibfnamefont {J.~F.}\ \bibnamefont {Morales}}, \bibinfo {author} {\bibfnamefont {P.}~\bibnamefont {Pani}}, \ and\ \bibinfo {author} {\bibfnamefont {G.}~\bibnamefont {Raposo}},\ }\href {\doibase 10.1103/PhysRevLett.125.221601} {\bibfield  {journal} {\bibinfo  {journal} {Phys. Rev. Lett.}\ }\textbf {\bibinfo {volume} {125}},\ \bibinfo {pages} {221601} (\bibinfo {year} {2020})},\ \Eprint {http://arxiv.org/abs/2007.01743} {arXiv:2007.01743 [hep-th]} \BibitemShut {NoStop}%
\bibitem [{\citenamefont {Bianchi}\ \emph {et~al.}(2021)\citenamefont {Bianchi}, \citenamefont {Consoli}, \citenamefont {Grillo}, \citenamefont {Morales}, \citenamefont {Pani},\ and\ \citenamefont {Raposo}}]{Bianchi:2020miz}%
  \BibitemOpen
  \bibfield  {author} {\bibinfo {author} {\bibfnamefont {M.}~\bibnamefont {Bianchi}}, \bibinfo {author} {\bibfnamefont {D.}~\bibnamefont {Consoli}}, \bibinfo {author} {\bibfnamefont {A.}~\bibnamefont {Grillo}}, \bibinfo {author} {\bibfnamefont {J.~F.}\ \bibnamefont {Morales}}, \bibinfo {author} {\bibfnamefont {P.}~\bibnamefont {Pani}}, \ and\ \bibinfo {author} {\bibfnamefont {G.}~\bibnamefont {Raposo}},\ }\href {\doibase 10.1007/JHEP01(2021)003} {\bibfield  {journal} {\bibinfo  {journal} {JHEP}\ }\textbf {\bibinfo {volume} {01}},\ \bibinfo {pages} {003} (\bibinfo {year} {2021})},\ \Eprint {http://arxiv.org/abs/2008.01445} {arXiv:2008.01445 [hep-th]} \BibitemShut {NoStop}%
\bibitem [{\citenamefont {Afshordi}\ \emph {et~al.}(2024)\citenamefont {Afshordi} \emph {et~al.}}]{Buoninfante:2024oxl}%
  \BibitemOpen
  \bibfield  {author} {\bibinfo {author} {\bibfnamefont {N.}~\bibnamefont {Afshordi}} \emph {et~al.}\ }(\bibinfo {year} {2024})\ \Eprint {http://arxiv.org/abs/2410.14414} {arXiv:2410.14414 [gr-qc]} \BibitemShut {NoStop}%
\bibitem [{\citenamefont {Carballo-Rubio}\ \emph {et~al.}(2025)\citenamefont {Carballo-Rubio} \emph {et~al.}}]{Carballo-Rubio:2025fnc}%
  \BibitemOpen
  \bibfield  {author} {\bibinfo {author} {\bibfnamefont {R.}~\bibnamefont {Carballo-Rubio}} \emph {et~al.},\ }\href@noop {} {\  (\bibinfo {year} {2025})},\ \Eprint {http://arxiv.org/abs/2501.05505} {arXiv:2501.05505 [gr-qc]} \BibitemShut {NoStop}%
\bibitem [{\citenamefont {Cardoso}\ and\ \citenamefont {Pani}(2017)}]{Cardoso:2017cqb}%
  \BibitemOpen
  \bibfield  {author} {\bibinfo {author} {\bibfnamefont {V.}~\bibnamefont {Cardoso}}\ and\ \bibinfo {author} {\bibfnamefont {P.}~\bibnamefont {Pani}},\ }\href {\doibase 10.1038/s41550-017-0225-y} {\bibfield  {journal} {\bibinfo  {journal} {Nature Astron.}\ }\textbf {\bibinfo {volume} {1}},\ \bibinfo {pages} {586} (\bibinfo {year} {2017})},\ \Eprint {http://arxiv.org/abs/1709.01525} {arXiv:1709.01525 [gr-qc]} \BibitemShut {NoStop}%
\bibitem [{\citenamefont {Mark}\ \emph {et~al.}(2017)\citenamefont {Mark}, \citenamefont {Zimmerman}, \citenamefont {Du},\ and\ \citenamefont {Chen}}]{Mark:2017dnq}%
  \BibitemOpen
  \bibfield  {author} {\bibinfo {author} {\bibfnamefont {Z.}~\bibnamefont {Mark}}, \bibinfo {author} {\bibfnamefont {A.}~\bibnamefont {Zimmerman}}, \bibinfo {author} {\bibfnamefont {S.~M.}\ \bibnamefont {Du}}, \ and\ \bibinfo {author} {\bibfnamefont {Y.}~\bibnamefont {Chen}},\ }\href {\doibase 10.1103/PhysRevD.96.084002} {\bibfield  {journal} {\bibinfo  {journal} {Phys. Rev. D}\ }\textbf {\bibinfo {volume} {96}},\ \bibinfo {pages} {084002} (\bibinfo {year} {2017})},\ \Eprint {http://arxiv.org/abs/1706.06155} {arXiv:1706.06155 [gr-qc]} \BibitemShut {NoStop}%
\bibitem [{\citenamefont {Carballo-Rubio}\ \emph {et~al.}(2018{\natexlab{b}})\citenamefont {Carballo-Rubio}, \citenamefont {Di~Filippo}, \citenamefont {Liberati},\ and\ \citenamefont {Visser}}]{Carballo-Rubio:2018jzw}%
  \BibitemOpen
  \bibfield  {author} {\bibinfo {author} {\bibfnamefont {R.}~\bibnamefont {Carballo-Rubio}}, \bibinfo {author} {\bibfnamefont {F.}~\bibnamefont {Di~Filippo}}, \bibinfo {author} {\bibfnamefont {S.}~\bibnamefont {Liberati}}, \ and\ \bibinfo {author} {\bibfnamefont {M.}~\bibnamefont {Visser}},\ }\href {\doibase 10.1103/PhysRevD.98.124009} {\bibfield  {journal} {\bibinfo  {journal} {Phys. Rev. D}\ }\textbf {\bibinfo {volume} {98}},\ \bibinfo {pages} {124009} (\bibinfo {year} {2018}{\natexlab{b}})},\ \Eprint {http://arxiv.org/abs/1809.08238} {arXiv:1809.08238 [gr-qc]} \BibitemShut {NoStop}%
\bibitem [{\citenamefont {Mazza}\ \emph {et~al.}(2021)\citenamefont {Mazza}, \citenamefont {Franzin},\ and\ \citenamefont {Liberati}}]{Mazza:2021rgq}%
  \BibitemOpen
  \bibfield  {author} {\bibinfo {author} {\bibfnamefont {J.}~\bibnamefont {Mazza}}, \bibinfo {author} {\bibfnamefont {E.}~\bibnamefont {Franzin}}, \ and\ \bibinfo {author} {\bibfnamefont {S.}~\bibnamefont {Liberati}},\ }\href {\doibase 10.1088/1475-7516/2021/04/082} {\bibfield  {journal} {\bibinfo  {journal} {JCAP}\ }\textbf {\bibinfo {volume} {04}},\ \bibinfo {pages} {082} (\bibinfo {year} {2021})},\ \Eprint {http://arxiv.org/abs/2102.01105} {arXiv:2102.01105 [gr-qc]} \BibitemShut {NoStop}%
\bibitem [{\citenamefont {Cardoso}\ and\ \citenamefont {Duque}(2022)}]{Cardoso:2022fbq}%
  \BibitemOpen
  \bibfield  {author} {\bibinfo {author} {\bibfnamefont {V.}~\bibnamefont {Cardoso}}\ and\ \bibinfo {author} {\bibfnamefont {F.}~\bibnamefont {Duque}},\ }\href {\doibase 10.1103/PhysRevD.105.104023} {\bibfield  {journal} {\bibinfo  {journal} {Phys. Rev. D}\ }\textbf {\bibinfo {volume} {105}},\ \bibinfo {pages} {104023} (\bibinfo {year} {2022})},\ \Eprint {http://arxiv.org/abs/2204.05315} {arXiv:2204.05315 [gr-qc]} \BibitemShut {NoStop}%
\bibitem [{\citenamefont {Casadio}\ \emph {et~al.}(2024)\citenamefont {Casadio}, \citenamefont {Kamenshchik},\ and\ \citenamefont {Ovalle}}]{Casadio:2024lgw}%
  \BibitemOpen
  \bibfield  {author} {\bibinfo {author} {\bibfnamefont {R.}~\bibnamefont {Casadio}}, \bibinfo {author} {\bibfnamefont {A.}~\bibnamefont {Kamenshchik}}, \ and\ \bibinfo {author} {\bibfnamefont {J.}~\bibnamefont {Ovalle}},\ }\href {\doibase 10.1103/PhysRevD.109.024042} {\bibfield  {journal} {\bibinfo  {journal} {Phys. Rev. D}\ }\textbf {\bibinfo {volume} {109}},\ \bibinfo {pages} {024042} (\bibinfo {year} {2024})},\ \Eprint {http://arxiv.org/abs/2401.03980} {arXiv:2401.03980 [gr-qc]} \BibitemShut {NoStop}%
\bibitem [{\citenamefont {Cardoso}\ and\ \citenamefont {Pani}(2019)}]{Cardoso:2019rvt}%
  \BibitemOpen
  \bibfield  {author} {\bibinfo {author} {\bibfnamefont {V.}~\bibnamefont {Cardoso}}\ and\ \bibinfo {author} {\bibfnamefont {P.}~\bibnamefont {Pani}},\ }\href {\doibase 10.1007/s41114-019-0020-4} {\bibfield  {journal} {\bibinfo  {journal} {Living Rev. Rel.}\ }\textbf {\bibinfo {volume} {22}},\ \bibinfo {pages} {4} (\bibinfo {year} {2019})},\ \Eprint {http://arxiv.org/abs/1904.05363} {arXiv:1904.05363 [gr-qc]} \BibitemShut {NoStop}%
\bibitem [{\citenamefont {Abedi}\ \emph {et~al.}(2017)\citenamefont {Abedi}, \citenamefont {Dykaar},\ and\ \citenamefont {Afshordi}}]{Abedi:2016hgu}%
  \BibitemOpen
  \bibfield  {author} {\bibinfo {author} {\bibfnamefont {J.}~\bibnamefont {Abedi}}, \bibinfo {author} {\bibfnamefont {H.}~\bibnamefont {Dykaar}}, \ and\ \bibinfo {author} {\bibfnamefont {N.}~\bibnamefont {Afshordi}},\ }\href {\doibase 10.1103/PhysRevD.96.082004} {\bibfield  {journal} {\bibinfo  {journal} {Phys. Rev. D}\ }\textbf {\bibinfo {volume} {96}},\ \bibinfo {pages} {082004} (\bibinfo {year} {2017})},\ \Eprint {http://arxiv.org/abs/1612.00266} {arXiv:1612.00266 [gr-qc]} \BibitemShut {NoStop}%
\bibitem [{\citenamefont {Jiang}\ \emph {et~al.}(2021)\citenamefont {Jiang}, \citenamefont {Wang}, \citenamefont {Yang},\ and\ \citenamefont {Wu}}]{Jiang:2021ajk}%
  \BibitemOpen
  \bibfield  {author} {\bibinfo {author} {\bibfnamefont {X.}~\bibnamefont {Jiang}}, \bibinfo {author} {\bibfnamefont {P.}~\bibnamefont {Wang}}, \bibinfo {author} {\bibfnamefont {H.}~\bibnamefont {Yang}}, \ and\ \bibinfo {author} {\bibfnamefont {H.}~\bibnamefont {Wu}},\ }\href {\doibase 10.1140/epjc/s10052-021-09816-z} {\bibfield  {journal} {\bibinfo  {journal} {Eur. Phys. J. C}\ }\textbf {\bibinfo {volume} {81}},\ \bibinfo {pages} {1043} (\bibinfo {year} {2021})},\ \bibinfo {note} {[Erratum: Eur.Phys.J.C 82, 5 (2022)]},\ \Eprint {http://arxiv.org/abs/2107.10758} {arXiv:2107.10758 [gr-qc]} \BibitemShut {NoStop}%
\bibitem [{\citenamefont {Shaikh}(2023)}]{Shaikh:2022ivr}%
  \BibitemOpen
  \bibfield  {author} {\bibinfo {author} {\bibfnamefont {R.}~\bibnamefont {Shaikh}},\ }\href {\doibase 10.1093/mnras/stad1383} {\bibfield  {journal} {\bibinfo  {journal} {Mon. Not. Roy. Astron. Soc.}\ }\textbf {\bibinfo {volume} {523}},\ \bibinfo {pages} {375} (\bibinfo {year} {2023})},\ \Eprint {http://arxiv.org/abs/2208.01995} {arXiv:2208.01995 [gr-qc]} \BibitemShut {NoStop}%
\bibitem [{\citenamefont {Bambi}\ \emph {et~al.}(2025)\citenamefont {Bambi} \emph {et~al.}}]{Bambi:2025wjx}%
  \BibitemOpen
  \bibfield  {author} {\bibinfo {author} {\bibfnamefont {C.}~\bibnamefont {Bambi}} \emph {et~al.}\ }(\bibinfo {year} {2025})\ \Eprint {http://arxiv.org/abs/2505.09014} {arXiv:2505.09014 [gr-qc]} \BibitemShut {NoStop}%
\bibitem [{\citenamefont {Ryan}(1995)}]{Ryan:1995wh}%
  \BibitemOpen
  \bibfield  {author} {\bibinfo {author} {\bibfnamefont {F.~D.}\ \bibnamefont {Ryan}},\ }\href {\doibase 10.1103/PhysRevD.52.5707} {\bibfield  {journal} {\bibinfo  {journal} {Phys. Rev. D}\ }\textbf {\bibinfo {volume} {52}},\ \bibinfo {pages} {5707} (\bibinfo {year} {1995})}\BibitemShut {NoStop}%
\bibitem [{\citenamefont {Pappas}\ and\ \citenamefont {Apostolatos}(2012)}]{Pappas:2012ns}%
  \BibitemOpen
  \bibfield  {author} {\bibinfo {author} {\bibfnamefont {G.}~\bibnamefont {Pappas}}\ and\ \bibinfo {author} {\bibfnamefont {T.~A.}\ \bibnamefont {Apostolatos}},\ }\href {\doibase 10.1103/PhysRevLett.108.231104} {\bibfield  {journal} {\bibinfo  {journal} {Phys. Rev. Lett.}\ }\textbf {\bibinfo {volume} {108}},\ \bibinfo {pages} {231104} (\bibinfo {year} {2012})},\ \Eprint {http://arxiv.org/abs/1201.6067} {arXiv:1201.6067 [gr-qc]} \BibitemShut {NoStop}%
\bibitem [{\citenamefont {Geroch}(1970{\natexlab{a}})}]{Geroch:1970cc}%
  \BibitemOpen
  \bibfield  {author} {\bibinfo {author} {\bibfnamefont {R.~P.}\ \bibnamefont {Geroch}},\ }\href {\doibase 10.1063/1.1665348} {\bibfield  {journal} {\bibinfo  {journal} {J. Math. Phys.}\ }\textbf {\bibinfo {volume} {11}},\ \bibinfo {pages} {1955} (\bibinfo {year} {1970}{\natexlab{a}})}\BibitemShut {NoStop}%
\bibitem [{\citenamefont {Geroch}(1970{\natexlab{b}})}]{Geroch:1970cd}%
  \BibitemOpen
  \bibfield  {author} {\bibinfo {author} {\bibfnamefont {R.~P.}\ \bibnamefont {Geroch}},\ }\href {\doibase 10.1063/1.1665427} {\bibfield  {journal} {\bibinfo  {journal} {J. Math. Phys.}\ }\textbf {\bibinfo {volume} {11}},\ \bibinfo {pages} {2580} (\bibinfo {year} {1970}{\natexlab{b}})}\BibitemShut {NoStop}%
\bibitem [{\citenamefont {Hansen}(1974)}]{Hansen:1974zz}%
  \BibitemOpen
  \bibfield  {author} {\bibinfo {author} {\bibfnamefont {R.~O.}\ \bibnamefont {Hansen}},\ }\href {\doibase 10.1063/1.1666501} {\bibfield  {journal} {\bibinfo  {journal} {J. Math. Phys.}\ }\textbf {\bibinfo {volume} {15}},\ \bibinfo {pages} {46} (\bibinfo {year} {1974})}\BibitemShut {NoStop}%
\bibitem [{\citenamefont {Thorne}(1980)}]{Thorne:1980ru}%
  \BibitemOpen
  \bibfield  {author} {\bibinfo {author} {\bibfnamefont {K.~S.}\ \bibnamefont {Thorne}},\ }\href {\doibase 10.1103/RevModPhys.52.299} {\bibfield  {journal} {\bibinfo  {journal} {Rev. Mod. Phys.}\ }\textbf {\bibinfo {volume} {52}},\ \bibinfo {pages} {299} (\bibinfo {year} {1980})}\BibitemShut {NoStop}%
\bibitem [{\citenamefont {{G{\"u}rsel}}(1983)}]{Gursel1983}%
  \BibitemOpen
  \bibfield  {author} {\bibinfo {author} {\bibfnamefont {Y.}~\bibnamefont {{G{\"u}rsel}}},\ }\href {\doibase 10.1007/BF01031881} {\bibfield  {journal} {\bibinfo  {journal} {General Relativity and Gravitation}\ }\textbf {\bibinfo {volume} {15}},\ \bibinfo {pages} {737} (\bibinfo {year} {1983})}\BibitemShut {NoStop}%
\bibitem [{\citenamefont {Friedman}(1978)}]{Friedman:1978ygc}%
  \BibitemOpen
  \bibfield  {author} {\bibinfo {author} {\bibfnamefont {J.~L.}\ \bibnamefont {Friedman}},\ }\href {\doibase 10.1007/BF01196933} {\bibfield  {journal} {\bibinfo  {journal} {Commun. Math. Phys.}\ }\textbf {\bibinfo {volume} {63}},\ \bibinfo {pages} {243} (\bibinfo {year} {1978})}\BibitemShut {NoStop}%
\bibitem [{\citenamefont {Cardoso}\ \emph {et~al.}(2008)\citenamefont {Cardoso}, \citenamefont {Pani}, \citenamefont {Cadoni},\ and\ \citenamefont {Cavaglia}}]{Cardoso:2007az}%
  \BibitemOpen
  \bibfield  {author} {\bibinfo {author} {\bibfnamefont {V.}~\bibnamefont {Cardoso}}, \bibinfo {author} {\bibfnamefont {P.}~\bibnamefont {Pani}}, \bibinfo {author} {\bibfnamefont {M.}~\bibnamefont {Cadoni}}, \ and\ \bibinfo {author} {\bibfnamefont {M.}~\bibnamefont {Cavaglia}},\ }\href {\doibase 10.1103/PhysRevD.77.124044} {\bibfield  {journal} {\bibinfo  {journal} {Phys. Rev. D}\ }\textbf {\bibinfo {volume} {77}},\ \bibinfo {pages} {124044} (\bibinfo {year} {2008})},\ \Eprint {http://arxiv.org/abs/0709.0532} {arXiv:0709.0532 [gr-qc]} \BibitemShut {NoStop}%
\bibitem [{\citenamefont {Moschidis}(2018)}]{Moschidis:2016zjy}%
  \BibitemOpen
  \bibfield  {author} {\bibinfo {author} {\bibfnamefont {G.}~\bibnamefont {Moschidis}},\ }\href {\doibase 10.1007/s00220-017-3010-y} {\bibfield  {journal} {\bibinfo  {journal} {Commun. Math. Phys.}\ }\textbf {\bibinfo {volume} {358}},\ \bibinfo {pages} {437} (\bibinfo {year} {2018})},\ \Eprint {http://arxiv.org/abs/1608.02035} {arXiv:1608.02035 [math.AP]} \BibitemShut {NoStop}%
\bibitem [{\citenamefont {Hawking}(1972)}]{Hawking:1971vc}%
  \BibitemOpen
  \bibfield  {author} {\bibinfo {author} {\bibfnamefont {S.~W.}\ \bibnamefont {Hawking}},\ }\href {\doibase 10.1007/BF01877517} {\bibfield  {journal} {\bibinfo  {journal} {Commun. Math. Phys.}\ }\textbf {\bibinfo {volume} {25}},\ \bibinfo {pages} {152} (\bibinfo {year} {1972})}\BibitemShut {NoStop}%
\bibitem [{\citenamefont {Hawking}\ and\ \citenamefont {Ellis}(2023)}]{Hawking:1973uf}%
  \BibitemOpen
  \bibfield  {author} {\bibinfo {author} {\bibfnamefont {S.~W.}\ \bibnamefont {Hawking}}\ and\ \bibinfo {author} {\bibfnamefont {G.~F.~R.}\ \bibnamefont {Ellis}},\ }\href {\doibase 10.1017/9781009253161} {\emph {\bibinfo {title} {{The Large Scale Structure of Space-Time}}}},\ Cambridge Monographs on Mathematical Physics\ (\bibinfo  {publisher} {Cambridge University Press},\ \bibinfo {year} {2023})\BibitemShut {NoStop}%
\bibitem [{\citenamefont {Robinson}(1975)}]{Robinson:1975bv}%
  \BibitemOpen
  \bibfield  {author} {\bibinfo {author} {\bibfnamefont {D.~C.}\ \bibnamefont {Robinson}},\ }\href {\doibase 10.1103/PhysRevLett.34.905} {\bibfield  {journal} {\bibinfo  {journal} {Phys. Rev. Lett.}\ }\textbf {\bibinfo {volume} {34}},\ \bibinfo {pages} {905} (\bibinfo {year} {1975})}\BibitemShut {NoStop}%
\bibitem [{\citenamefont {Emparan}(2025)}]{Emparan:2025wsh}%
  \BibitemOpen
  \bibfield  {author} {\bibinfo {author} {\bibfnamefont {R.}~\bibnamefont {Emparan}},\ }\href {\doibase 10.1007/s10714-025-03398-x} {\bibfield  {journal} {\bibinfo  {journal} {Gen. Rel. Grav.}\ }\textbf {\bibinfo {volume} {57}},\ \bibinfo {pages} {73} (\bibinfo {year} {2025})}\BibitemShut {NoStop}%
\bibitem [{\citenamefont {Bianchi}\ \emph {et~al.}(2024)\citenamefont {Bianchi}, \citenamefont {Gambino}, \citenamefont {Pani},\ and\ \citenamefont {Riccioni}}]{Bianchi:2024shc}%
  \BibitemOpen
  \bibfield  {author} {\bibinfo {author} {\bibfnamefont {M.}~\bibnamefont {Bianchi}}, \bibinfo {author} {\bibfnamefont {C.}~\bibnamefont {Gambino}}, \bibinfo {author} {\bibfnamefont {P.}~\bibnamefont {Pani}}, \ and\ \bibinfo {author} {\bibfnamefont {F.}~\bibnamefont {Riccioni}},\ }\href@noop {} {\  (\bibinfo {year} {2024})},\ \Eprint {http://arxiv.org/abs/2412.01771} {arXiv:2412.01771 [gr-qc]} \BibitemShut {NoStop}%
\bibitem [{\citenamefont {Bonga}\ and\ \citenamefont {Yang}(2021)}]{Bonga:2021ouq}%
  \BibitemOpen
  \bibfield  {author} {\bibinfo {author} {\bibfnamefont {B.}~\bibnamefont {Bonga}}\ and\ \bibinfo {author} {\bibfnamefont {H.}~\bibnamefont {Yang}},\ }\href {\doibase 10.1103/PhysRevD.104.084040} {\bibfield  {journal} {\bibinfo  {journal} {Phys. Rev. D}\ }\textbf {\bibinfo {volume} {104}},\ \bibinfo {pages} {084040} (\bibinfo {year} {2021})},\ \Eprint {http://arxiv.org/abs/2106.08342} {arXiv:2106.08342 [gr-qc]} \BibitemShut {NoStop}%
\bibitem [{\citenamefont {Gambino}\ \emph {et~al.}(2024)\citenamefont {Gambino}, \citenamefont {Pani},\ and\ \citenamefont {Riccioni}}]{Gambino:2024uge}%
  \BibitemOpen
  \bibfield  {author} {\bibinfo {author} {\bibfnamefont {C.}~\bibnamefont {Gambino}}, \bibinfo {author} {\bibfnamefont {P.}~\bibnamefont {Pani}}, \ and\ \bibinfo {author} {\bibfnamefont {F.}~\bibnamefont {Riccioni}},\ }\href {\doibase 10.1103/PhysRevD.109.124018} {\bibfield  {journal} {\bibinfo  {journal} {Phys. Rev. D}\ }\textbf {\bibinfo {volume} {109}},\ \bibinfo {pages} {124018} (\bibinfo {year} {2024})},\ \Eprint {http://arxiv.org/abs/2403.16574} {arXiv:2403.16574 [hep-th]} \BibitemShut {NoStop}%
\bibitem [{\citenamefont {Amalberti}\ \emph {et~al.}(2024)\citenamefont {Amalberti}, \citenamefont {Larrouturou},\ and\ \citenamefont {Yang}}]{Amalberti:2023ohj}%
  \BibitemOpen
  \bibfield  {author} {\bibinfo {author} {\bibfnamefont {L.}~\bibnamefont {Amalberti}}, \bibinfo {author} {\bibfnamefont {F.}~\bibnamefont {Larrouturou}}, \ and\ \bibinfo {author} {\bibfnamefont {Z.}~\bibnamefont {Yang}},\ }\href {\doibase 10.1103/PhysRevD.109.104027} {\bibfield  {journal} {\bibinfo  {journal} {Phys. Rev. D}\ }\textbf {\bibinfo {volume} {109}},\ \bibinfo {pages} {104027} (\bibinfo {year} {2024})},\ \Eprint {http://arxiv.org/abs/2312.02868} {arXiv:2312.02868 [gr-qc]} \BibitemShut {NoStop}%
\bibitem [{\citenamefont {Heynen}\ and\ \citenamefont {Mayerson}(2023)}]{Heynen:2023sin}%
  \BibitemOpen
  \bibfield  {author} {\bibinfo {author} {\bibfnamefont {J.}~\bibnamefont {Heynen}}\ and\ \bibinfo {author} {\bibfnamefont {D.~R.}\ \bibnamefont {Mayerson}},\ }\href@noop {} {\  (\bibinfo {year} {2023})},\ \Eprint {http://arxiv.org/abs/2312.04352} {arXiv:2312.04352 [gr-qc]} \BibitemShut {NoStop}%
\bibitem [{\citenamefont {Nicolini}\ \emph {et~al.}(2006)\citenamefont {Nicolini}, \citenamefont {Smailagic},\ and\ \citenamefont {Spallucci}}]{Nicolini:2005vd}%
  \BibitemOpen
  \bibfield  {author} {\bibinfo {author} {\bibfnamefont {P.}~\bibnamefont {Nicolini}}, \bibinfo {author} {\bibfnamefont {A.}~\bibnamefont {Smailagic}}, \ and\ \bibinfo {author} {\bibfnamefont {E.}~\bibnamefont {Spallucci}},\ }\href {\doibase 10.1016/j.physletb.2005.11.004} {\bibfield  {journal} {\bibinfo  {journal} {Phys. Lett. B}\ }\textbf {\bibinfo {volume} {632}},\ \bibinfo {pages} {547} (\bibinfo {year} {2006})},\ \Eprint {http://arxiv.org/abs/gr-qc/0510112} {arXiv:gr-qc/0510112} \BibitemShut {NoStop}%
\bibitem [{\citenamefont {Israel}(1970)}]{Israel:1970kp}%
  \BibitemOpen
  \bibfield  {author} {\bibinfo {author} {\bibfnamefont {W.}~\bibnamefont {Israel}},\ }\href {\doibase 10.1103/PhysRevD.2.641} {\bibfield  {journal} {\bibinfo  {journal} {Phys. Rev. D}\ }\textbf {\bibinfo {volume} {2}},\ \bibinfo {pages} {641} (\bibinfo {year} {1970})}\BibitemShut {NoStop}%
\bibitem [{\citenamefont {Balasin}\ and\ \citenamefont {Nachbagauer}(1994)}]{Balasin:1993kf}%
  \BibitemOpen
  \bibfield  {author} {\bibinfo {author} {\bibfnamefont {H.}~\bibnamefont {Balasin}}\ and\ \bibinfo {author} {\bibfnamefont {H.}~\bibnamefont {Nachbagauer}},\ }\href {\doibase 10.1088/0264-9381/11/6/010} {\bibfield  {journal} {\bibinfo  {journal} {Class. Quant. Grav.}\ }\textbf {\bibinfo {volume} {11}},\ \bibinfo {pages} {1453} (\bibinfo {year} {1994})},\ \Eprint {http://arxiv.org/abs/gr-qc/9312028} {arXiv:gr-qc/9312028} \BibitemShut {NoStop}%
\bibitem [{\citenamefont {Myers}\ and\ \citenamefont {Perry}(1986)}]{Myers:1986un}%
  \BibitemOpen
  \bibfield  {author} {\bibinfo {author} {\bibfnamefont {R.~C.}\ \bibnamefont {Myers}}\ and\ \bibinfo {author} {\bibfnamefont {M.~J.}\ \bibnamefont {Perry}},\ }\href {\doibase 10.1016/0003-4916(86)90186-7} {\bibfield  {journal} {\bibinfo  {journal} {Annals Phys.}\ }\textbf {\bibinfo {volume} {172}},\ \bibinfo {pages} {304} (\bibinfo {year} {1986})}\BibitemShut {NoStop}%
\bibitem [{\citenamefont {Rezzolla}\ and\ \citenamefont {Zanotti}(2013)}]{Rezzolla:2013dea}%
  \BibitemOpen
  \bibfield  {author} {\bibinfo {author} {\bibfnamefont {L.}~\bibnamefont {Rezzolla}}\ and\ \bibinfo {author} {\bibfnamefont {O.}~\bibnamefont {Zanotti}},\ }\href {\doibase 10.1093/acprof:oso/9780198528906.001.0001} {\emph {\bibinfo {title} {{Relativistic Hydrodynamics}}}}\ (\bibinfo  {publisher} {Oxford University Press},\ \bibinfo {year} {2013})\BibitemShut {NoStop}%
\bibitem [{\citenamefont {Romatschke}\ and\ \citenamefont {Romatschke}(2019)}]{Romatschke:2017ejr}%
  \BibitemOpen
  \bibfield  {author} {\bibinfo {author} {\bibfnamefont {P.}~\bibnamefont {Romatschke}}\ and\ \bibinfo {author} {\bibfnamefont {U.}~\bibnamefont {Romatschke}},\ }\href {\doibase 10.1017/9781108651998} {\emph {\bibinfo {title} {{Relativistic Fluid Dynamics In and Out of Equilibrium}}}},\ Cambridge Monographs on Mathematical Physics\ (\bibinfo  {publisher} {Cambridge University Press},\ \bibinfo {year} {2019})\ \Eprint {http://arxiv.org/abs/1712.05815} {arXiv:1712.05815 [nucl-th]} \BibitemShut {NoStop}%
\bibitem [{\citenamefont {Gradshteyn}\ and\ \citenamefont {Ryzhik}(1943)}]{Gradshteyn:1943cpj}%
  \BibitemOpen
  \bibfield  {author} {\bibinfo {author} {\bibfnamefont {I.~S.}\ \bibnamefont {Gradshteyn}}\ and\ \bibinfo {author} {\bibfnamefont {I.~M.}\ \bibnamefont {Ryzhik}},\ }\href@noop {} {\emph {\bibinfo {title} {{Table of Integrals, Series, and Products}}}}\ (\bibinfo {year} {1943})\BibitemShut {NoStop}%
\end{thebibliography}%

\end{document}